\documentclass[aps,prb,preprint,groupedaddress,showpacs,tightenlines]{revtex4-1}

\usepackage{graphicx}%
\usepackage{longtable}

\begin{document}


\title{Gallium interstitial contributions to diffusion in gallium arsenide}


\author{J. T. Schick}
\email[]{joseph.schick@villanova.edu}
\affiliation{Physics Department, Villanova University,
Villanova, Pennsylvania 19085}

\author{C. G. Morgan}
\author{P. Papoulias}
\affiliation{Department of Physics and Astronomy, Wayne State University,
Detroit, Michigan 48202}


\date{\today}

\begin{abstract}
Enthalpies of formation of gallium interstitials and all the other native 
point defects in gallium arsenide are calculated using the same 
well-converged \emph{ab initio} techniques.  Using these results, equilibrium 
concentrations of these defects are computed as a function of chemical 
potential from the arsenic rich limit to the gallium rich limit and 
as a function of the doping level from $p$-type to $n$-type.  Gallium 
interstitial diffusion paths and migration barriers for diffusion are 
determined for all the interstitial charge states which are favored for 
Fermi levels anywhere in the gap, and the charge states which dominate 
diffusion as a function of Fermi level are identified.  The effects of chemical 
potential, doping level, and non-equilibrium defect concentrations 
produced by ion implantation or irradiation on gallium self-diffusion 
are examined.  Results are consistent with experimental results across 
the ranges of doping and stoichometry where comparisons can be 
made.
Finally, these calculations shed some light on the complex situation
for gallium diffusion in gallium arsenide that is gallium-rich and
doped heavily $p$-type.
\end{abstract}


\maketitle


\section{\label{section:introduction}Introduction}

Gallium arsenide is used in numerous applications in which converting
between electrical energy and light is important, and is therefore an
important material of continuing technological interest.
The successful operation of gallium-arsenide-based devices
generally depends on their structure of layers
of varying composition and/or doping, and the interfaces between these layers.
The motions of native, iso-electronic substituent, and dopant atoms are
manipulated during growth and fabrication to produce the desired atomic
composition in each region.  
After fabrication, the potential for continued 
movement of these atoms must be understood and controlled to avoid
degradation of performance and eventual device failure.

Native defects such as interstitials, vacancies, and antisites are often electrically active,
and can strongly affect device characteristics and longevity.  Motion of native point 
defects, such as vacancies and interstitials produced by non-stoichiometric growth 
or irradiation at the surface, can lead to formation of voids, interstitial clusters, 
precipitates, and other extended defects, as well as altered spatial profiles of the 
point defects.  The processes involving motion of the native point defects which 
control self-diffusion are often important for diffusion of foreign atoms as well.

The diffusion properties of a particular native defect are determined by
 the energy barriers for  migration of that defect
along its potential pathways.  When these energy barriers are combined with
the concentration of the defect in the material and the potential 
for interaction with foreign atoms, the likelihood of this
particular defect to enhance both self-diffusion and the diffusion of the foreign atoms
can be estimated.
Native defects outside the crystal lattice, such as interstitials, 
generally possess lower migration barriers than defects on the lattice, 
such as vacancies.\cite{Mehrer2007book}
Although much experimental and theoretical work has been done to 
determine and characterize the most important microscopic mechanisms 
for gallium interstitial diffusion in gallium
arsenide, there are still gaps in
the current overall picture of gallium interstitial diffusion in this material.

Self-diffusion is more difficult to examine experimentally than foreign atom
diffusion because concentrations of the native species may be
nearly uniform across the sample, and looking at these concentrations cannot identify
the individual motions of the identical atoms.  Calculations fit to the
experimental results for the simultaneous diffusion of two species, for
example interdiffusion of gallium and aluminum at an interface between
gallium arsenide and aluminum gallium arsenide, require that a
reasonably simple model with a small number of fitting parameters
is sufficient to describe the experimental situation.  However, studies
of the broadening of boundary layers in
multi-layered structures of gallium arsenide 
and aluminum gallium arsenide have been extensively used to
investigate diffusion of gallium and aluminum in
Al$_x$Ga$_{1-x}$As.\cite{Mehrer2007book,Tan1991a}
More direct methods for determining diffusion barriers
for self-diffusion may use material grown or implanted with an isotope tracer in one region.
After annealing and subsequent measurement of the spatial dependence
of the isotope concentrations as a function of annealing time, temperature, and
Fermi level, theoretical model calculations
are fitted to the observed concentration profiles,\cite{Mehrer2007book}
and these fits are used to generate numerical values for the formation 
and migration energies of the defects that are assumed to dominate
the diffusion under the existing experimental conditions. 
In some cases, one model is proposed and shown to fit the experimental results,\cite{Bracht2005a} 
and in other cases, more than one alternative model is found which can fit the experimental results.\cite{Koumetz2006a}

Initially, gallium interstitial diffusion was proposed to be modeled
well with doubly or triply positive charge states.\cite{Tan1991a}
This proposal was supported by early analyses of diffusion experiments 
involving zinc and cadmium, which are both known to be substitutional 
$p$-type dopants in the gallium sublattice.\cite{Tan1991a}
Zucker \emph{et al.}\ studied the effects of ion-implanted zinc on the
boundary profiles of superlattices of gallium arsenide and aluminum
gallium arsenide and drew the conclusion that gallium interstitials
were diffusing in a doubly or triply positive charge state.\cite{Zucker1989a}
B{\"o}sker \emph{et al.}\ diffused zinc into gallium arsenide under
arsenic rich conditions\cite{Bosker1995a} and cadmium into gallium
arsenide under arsenic rich and arsenic poor conditions,\cite{Bosker1999a}
and also concluded that the doubly and triply positive charge states
of gallium interstitials can adequately model the diffusion 
profiles.\cite{Bosker1995a,Bosker1999a}

However, in a later study employing isotope heterostructures
with an implanted dopant to determine the
relationship between zinc and gallium diffusion
in gallium arsenide\cite{Bracht2001a} it was proposed 
that it is the neutral charge state of the gallium interstitial
and the neutral and singly-positive charge states of the
gallium vacancy that are important
in zinc-gallium co-diffusion, contrasting with the earlier
studies on interstitials.
In subsequent work, Bracht \emph{et al.}\cite{Bracht2005a}
examined zinc and gallium co-diffusion into the surface of
gallium arsenide, further refining their
conclusion that the diffusion profiles can be fitted successfully by requiring
 that the gallium interstitials which contribute to diffusion are
in a neutral or singly-positive charge state.  It should be
noted that using a zinc gallium alloy as a source for the indiffusing 
atoms in these experiments means
the material is being prepared both gallium rich and
$p$-type.  However, the authors state that only samples with
sufficiently low levels of zinc at the surface are kept for analysis.\cite{Bracht2005a}
This has the effect of avoiding heavy $p$-type doping.
The authors in 
Refs.\ \onlinecite{Bracht2005a} and \onlinecite{Bracht2001a} conclude
from their own measurements and from re-analysis of
the results of B{\"o}sker\cite{Bosker1995a,Bosker1999a} that 
interstitial gallium in the neutral and singly positive charge
states are partial contributors to gallium diffusion, with
gallium vacancies contributing more strongly over much of the
experimentally accessible range of stoichometry and doping level.  
Under arsenic-rich conditions,
these authors observe an enhanced contribution of
gallium vacancies to gallium diffusion.  Finally,
under gallium-rich conditions in heavily-$p$-doped material, the
authors have yet to determine a model that properly fits
the diffusion profile.\cite{Bracht2005a,BrachtPrivate}  This complex situation
requires a comprehensive theoretical picture capable of addressing
the numerous defect configurations and charge states that may contribute significantly
to the diffusion.


In an early computational effort that included a wide 
range of potentially important defects in gallium arsenide,
Zhang and Northrup used first-principles methods to calculate the formation energies
for antisites, tetrahedral interstitials, and vacancies of the native atoms 
as a function of doping from $p$-type to $n$-type
and as a function of chemical potential across the range
from arsenic-rich to gallium-rich.\cite{Zhang1991}
They concluded that gallium self-diffusion
in $n$-type arsenic-rich gallium arsenide would be dominated
by gallium vacancies, while in the $p$-type gallium-rich
limit, gallium interstitials in a highly-positive charge state
would dominate diffusion.\cite{Zhang1991}
Since the time of this study, 
computing power has substantially
increased, relaxing restrictions on
the size of supercells and the accuracy of approximations for the sum over all occupied 
states which can be achieved, thus allowing more 
accurate calculations of the properties of isolated point defects. Advances 
in computer power and program development have also 
expanded the number of possible starting
configurations that can be studied, and enhanced theorists' ability 
to explore the potential energy surface in the vicinity of these configurations.
It has been shown that lower-symmetry interstitials often have comparable or 
lower energies of formation than high-symmetry, tetrahedral 
interstitials.\cite{Chadi1992b}  For example, $\langle 110 \rangle$ split interstitials 
have a lower energy of formation than tetrahedral interstitials both for arsenic
interstitials in gallium arsenide\cite{Chadi1992,Landman1997,Schick2002a}
and for silicon interstitials in silicon
\cite{Bar-Yam1984a,Pantelides1983a,Needs1999a,Goedecker2002a}
over a significant range of Fermi levels in the gap.

In addition to the exploration of lower symmetry defects, several thorough 
convergence studies for defect calculations have also been 
performed.\cite{Puska1998,Schick2002a,Schultz2009a}  For example, it was shown that the 
smaller supercells and smaller number of $k$-points in sums over
$k$-space which were used in earlier calculations result in errors ranging 
from several tenths of eV to well over an eV in calculations of the defect 
formation and ionization energies for arsenic interstitials in
gallium arsenide.\cite{Schick2002a}  
Later calculations 
using the larger supercells and larger number of $k$-points which are required 
for convergence have concluded that tetrahedral interstitial configurations are 
the most energetically favorable for gallium interstitials in gallium arsenide, and 
have also investigated lower symetry configurations 
which may be important intermediate points along low-energy diffusion
paths.\cite{Malouin2007a}
Most recently, all the simple intrinsic defects in gallium arsenide have been
examined with a single approach, including the most thorough study to date of 
convergence and the effects of using different density functionals 
(local density and generalized gradient approximation).\cite{Schultz2009a}

Following these recent advances in calculational accuracy and the ability 
to consider diffusion paths without restriction to paths of high symmetry,
\emph{ab initio} studies
of the diffusion of gallium and arsenic
vacancies\cite{Bockstedte1997b,ElMellouhi2005a,ElMellouhi2006a}
and arsenic interstitials\cite{Papoulias2010,*Papoulias2009} 
in gallium arsenide have been carried out.
In recent \emph{ab initio} studies of the diffusion of gallium interstitials in gallium arsenide,
\cite{Levasseur-Smith2008a,Levasseur-Smith2008b} the migration 
barriers for gallium interstitials in gallium arsenide in the
neutral and singly positive charge states
have been calculated, and the resulting computed migration activation enthalpies
for interstitials in these two charge states were found to be comparable to the migration
activation enthalpies obtained from recent models fit to experiments.\cite{Bracht2005a}  
These computed migration activation enthalpies for
gallium interstitials in the neutral and singly positive charge
states\cite{Levasseur-Smith2008a} were actually 0.5 eV
lower than the
experimental values\cite{Bracht2005a} to which they were compared, 
which as the authors noted 
may be due to a number of reasons, ranging from the finite scaling 
correction used in the theory (for non-zero
charge state migrations) and the lack of thermal or entropic effects
in the theoretical calculations to 
assumptions made in the model calculations used for analysis of the experiment 
and assumptions about the material preparation used when 
comparing the theoretical 
to the experimental results.  However, the migration paths and activation enthalpies 
for the +2 and +3 charge states 
were not calculated, and no direct comparison between all the possible  
charge states was made to 
determine which charge states dominate gallium interstitial diffusion 
as a function of the Fermi level.  
In addition, in more recent work, Schultz {\em et al.}\cite{Schultz2009a} have determined that 
the last electron added to such density functional calculations to represent the 
neutral state of the tetrahedral gallium interstitial is 
not bound in a localized deep level, but occupies a state made out of 
conduction band edge states.  Schultz {\em et al.}\cite{Schultz2009a} conclude that any 
identification of favorable activation energies for neutral gallium 
interstitial diffusion based on calculations of the formation energy of neutral tetrahedral 
interstitials which assume that the last electron is bound to the interstitial in a localized 
deep level are not to be relied upon.

Because of the still-existing uncertainties
in analyses based upon both experimental and theoretical results,
a comprehensive picture of gallium interstitial diffusion, carefully 
considering the contributions of all possible charge states, is needed.   
In this paper we compute the
formation energies for all the native point defects in
gallium arsenide across the full range of allowed chemical potentials,
from the arsenic-rich limit to the gallium-rich limit,
and across the range of possible doping levels,
from $p$-type to $n$-type, using an \emph{ab initio} approach, with a single set
of approximations.  
We compute barriers for gallium interstitial
diffusion on low-energy pathways and compare our numbers to
all recently available calculations for gallium interstitial formation and diffusion.
We examine activation energies associated with diffusion across the full
ranges of chemical potential and doping.  Because gallium vacancies
will also contribute to diffusion of gallium, we also include in our analysis
the energies associated with gallium vacancy diffusion across the full
ranges of chemical potential and doping.  From these energetic
studies a comprehensive picture of gallium diffusion across experimentally
accessible conditions emerges.

\section{\label{section:approach}Approach}

For this study we utilized
codes\cite{Bockstedte1997,vasp1,vasp2,vasp3,vasp4}
implementing density-functional theory \cite{Hohenberg1964} within
the local density approximation, with the Ceperley-Alder
\cite{Ceperley1980} form for the exchange and correlation potentials
as parameterized by Perdew and Zunger.\cite{Perdew1981} The core
electrons are treated in the frozen-core approximation and the ion
cores are replaced by fully-separable \cite{Kleinman1982}
norm-conserving pseudopotentials.\cite{Hamann1989}
Plane waves are
included up to the energy cutoff of 10~Ry ($\approx 1.4 \times 10^{2}$~eV).
The atoms are allowed to
relax until the force components are are less than $5 \times
10^{-4}$~hartrees per bohr radius ($\approx 2.6 \times 10^{-2}$~eV/{\AA })
and the zero temperature formation
energies change by less than $5 \times 10^{-6}$\,hartrees
($\approx 1.4 \times 10^{-4}$~eV)
per step for at least 100 steps.

For calculations of gallium interstitial migration barriers,
we employed the \textsc{vasp}
code\cite{vasp1,vasp2, vasp3,vasp4} with
ultra-soft pseudopotentials\cite{vanderbilt1} as supplied
by G. Kresse and J. Hafner.\cite{vanderbilt2} 
The nudged elastic band method\cite{Jonsson1998a,Mills1995a} was
applied within \textsc{vasp} in order to determine the
lowest-energy pathways for the migration of defects.
All calculations were tested for consistency with the results from calculations
with norm-conserving pseudopotentials.  We started with the structural results of the
earlier calculations and relaxed the structure with the \textsc{vasp} code, imposing
the same convergence criteria as above, and found that the arrangements
of the atoms remained unchanged.
Energies computed with ultrasoft pseudopotentials
were uniformly about 0.5~eV lower than those computed with norm-conserving
pseudopotentials for all calculations involving a single interstitial gallium
atom.  The near-uniformity of this difference gives us confidence in the calculated 
results for the energy differences between different configurations, 
as well as use of the \textsc{vasp} code for determining the low 
energy migration paths and corresponding energy barriers.

The supercells we used in these calculations are based on the bulk, 216-atom
cubic supercell with the bulk lattice constant determined
through fitting the Murgnahan\cite{Murnaghan1944a} equation of state to
the calculated energies as a function of the volume of the primitive cell.
We found a lattice constant of 5.55~{\AA }, which is 1.7{\% } smaller
than the experimental value of 5.65~{\AA } at 300K and 2.2{\% } smaller
than the 1000K experimental value of
5.68~{\AA}.\cite{Blakemore1982a}
In those calculations in which we employed the \textsc{vasp} code we
used a lattice constant of 5.59~{\AA} determined by the same procedure.
(The \textsc{vasp} value is 1{\%} smaller than the 300K experimental value
and 1.6{\% } smaller than the 1000K experimental value of the lattice
constant.)\cite{Blakemore1982a}

In examination of the geometries of the defects, bonding characteristics,
and charge distributions, we have employed the software
\textsc{vesta}\cite{VESTA} to produce the necessary volumetric plots
from the results of our calculations.
The volumetric plots found below also have been produced through the
use of this program.

The free energy formalism\cite{Zhang1991} we adopt expresses the Gibbs free energy
of the system in terms of the chemical potentials of 
gallium ($\mu_{\mathrm{Ga,env}}$) and arsenic ($\mu_\mathrm{As,env}$) in the
environment.  For GaAs to be in equilibrium with its environment, these 
chemical potentials must be related to the chemical
potential per gallium-arsenic pair in the gallium arsenide crystal as follows: $\mu_\mathrm{GaAs}
= \mu_\mathrm{Ga,env} + \mu_\mathrm{As,env}$.  Furthermore, for gallium arsenide
to be stable, the magnitude of the difference between the chemical potentials of
gallium and arsenic in the environment is limited to a range
determined by the heat of formation of bulk gallium arsenide per atomic 
pair, $\mu_\mathrm{GaAs} - \mu_\mathrm{Ga}
- \mu_\mathrm{As}$, where $\mu_\mathrm{Ga}$ and  $\mu_\mathrm{As}$ are
the chemical potentials of a gallium atom in bulk gallium and an 
arsenic atom in bulk arsenic, respectively.
For convenient calculation, the dependence of the Gibbs free energy on
the chemical potential can be expressed in terms of the bulk chemical
potentials and a parameter $\Delta \mu \equiv (\mu_\mathrm{Ga,env} - \mu_\mathrm{As,env})
- (\mu_\mathrm{Ga} - \mu_\mathrm{As})$.  The stable range for gallium arsenide
material is determined by $\left| \Delta\mu \right| \le - \left( \mu_\mathrm{GaAs}
- \mu_\mathrm{Ga} - \mu_\mathrm{As} \right)$.
The resulting expression for the Gibbs free energy in this formalism\cite{Zhang1991} is
\begin{eqnarray}
\label{eq:gibbs}
G_f = E( N_{\mathrm{Ga}} , N_{\mathrm{As}} , N_{\mathrm{e}} )
     - TS + PV \nonumber \\
     - \frac{ N_{\mathrm{Ga}} + N_{\mathrm{As}} }{2}
         \mu_{\mathrm{GaAs}}  \nonumber \\
     - \frac{ N_{\mathrm{Ga}} - N_{\mathrm{As}} }{2}
         \left(\mu_{\mathrm{Ga}} - \mu_{\mathrm{As}} \right)
     -   \frac{ N_{\mathrm{Ga}} - N_{\mathrm{As}} }{2}
                \Delta\mu  \nonumber \\
     - N_{\mathrm{e}}\, \epsilon_F - N_{\mathrm{e}}\, \Delta \Phi
.
\end{eqnarray}
where $E(N_{\mathrm{Ga}}, N_{\mathrm{As}}, N_{\mathrm{e}})$ is the internal
energy calculated for a given defect in a supercell consisting of
$N_{\mathrm{Ga}}$ gallium atoms and $N_{\mathrm{As}}$ arsenic atoms
and possessing $N_{\mathrm{e}}$ (excess) electrons which have been transferred from a
reservoir with Fermi level $\epsilon_F$.  (The system is in charge state
$q = -N_{\mathrm{e}}$, and for defects involving a net excess of 
gallium or arsenic atoms, $N_{\mathrm{Ga}} - N_{\mathrm{As}}$ will be non-zero.)

In the gallium-rich (arsenic-rich) case, the defect is assumed to
form in a sample that is in equilibrium with pure gallium
(pure arsenic) with $\Delta \mu$ at its maximum (minimum) value.
All the chemical potentials needed for this calculation have been
computed using the same methods described above with norm-conserving
pseudopotentials.

 The formation entropy $S$
consists of contributions due to the multiplicity of configurations
related by symmetry operations, including different split interstitial 
orientations and Jahn-Teller distortions, and vibration.
The vibrational entropy will be the most significant and is prohibitively
expensive to calculate so it is not included in the results below.  
(The enthalpies of formation are used in the remaining
discussions in this paper.)
The $PV$ term is unimportant in the current calculation because 
the energy contribution per atom (or atom pair) of this term for the materials we are
discussing is on the order of $10^{-5}$~eV.\cite{Schick2002a}

Because there is no absolute alignment
of eigenstates between different calculations, we calculate the
electric potential along a line from the defect across the supercell
and averaged over planes perpendicular to this line, according to
the procedure outlined by Kohan {\em et al.}\cite{Kohan2000} Far
from the neutral defect the plane-averaged potential converges to a
fixed value.  From this fixed value, we subtract the plane-averaged
electric potential computed in the bulk for the same geometry to
obtain the potential difference $\Delta \Phi$.  This potential
difference causes the energy of a defect in charge state
$q = -N_{\mathrm{e}}$ to be artificially shifted by
$N_{\mathrm{e}} \, \Delta \Phi$ and we compensate by
subtracting this from the computed energies. (Typical values of $|
\Delta \Phi |$ are around 0.05~eV.)

Previously we demonstrated\cite{Schick2002a} that supercells smaller 
than 216 atoms do not give well-converged formation energies for isolated charged
arsenic interstitials, and therefore do not produce well-converged
charge transition energies for isolated arsenic interstitials in gallium
arsenide.
In this work we used 216-atom supercells (plus or minus defect atoms)
to produce the total energies of charged and neutral defects in gallium arsenide,
which are displayed in Figs.\ \ref{figure:enthalpy} and \ref{figure:asrichenthalpy}
for the gallium-rich and arsenic-rich limits, respectively.

The supercell approximation is known to include a non-physical interaction
between the artificially-duplicated defects in different supercells.
While an actual infinite-size (or even a very large-size) sample is 
out of the question, a suitable extrapolation may be
applied to obtain a better estimate of the formation energy in the large-supercell limit.
In the supercell calculation, the charge state of the defect is produced
by adding or removing electrons from the defect atom and placing
a homogeneous compensating charge in the background.
The redistribution of charge that results when self-consistency is obtained
will have a non-zero electrostatic energy of interaction between supercells,
which was suggested to be properly
removed by computing the dipole and quadrupole moments
of the supercell and then removing the energy of the interaction
between these artificially-repeated moments from the supercell
result.\cite{Makov1995a}
Generally, corrections of this type are more important for
defect charge states farther from neutral and for smaller
supercells.
In the case of the gallium interstitial in
the +3 state (the lowest energy state when
the Fermi energy is taken at the
valence band edge) we evaluated these corrections
and found that they are
$+2.408$~eV for the 65-atom supercell and 
$-0.287$~eV for the 217-atom supercell.
The uncorrected energy in the larger cell is 0.3~eV higher
than in the smaller cell.
However adding the correction
\emph{increases} the difference between the smaller and larger supercells to
around 2.4~eV in the opposite sense to the uncorrected difference.
The lack of convergence in the results leads us to doubt
the validity of this correction in this situation.

In a related approach to this correction, making note of the fact
the original correction depends upon the bulk dielectric
constant, an external parameter, some investigators
fit the results using the expression 
$AL^{-1}+BL^{-3}$.  This expression scales with supercell lattice parameter $L$ 
exactly as the Makov model does, but treats the interrelated
pre-factors as fitting parameters to `data' constituting of formation energies 
computed for different $L$ values.
Corrections of this form have been
employed in a variety calculations, and yet
they remain controversial.
Work in this area\cite{Vandewalle2004a} has shown 
that this widely-applied finite scaling approach,\cite{Makov1995a}
while certainly accounting for the basic effect, does not consistently
produce the infinite-size limiting result.
Recently a new, but related, approach has been outlined
that may be used to properly remove
the supercell size effect.\cite{freysoldt2009a}
For a discussion of the situation surrounding the use of these corrections,
see the review by Nieminen.\cite{Niemenen2007a}
As a result of the continuing discussion and the uncertainty introduced
by the numbers in our test calculations,
we have not employed these corrections in the present results.

\section{\label{section:resultsanddiscussion}Results \& Discussion}

\subsection{\label{subsection:energyandstructure}Defect energy and structure}

\begin{figure}
\includegraphics{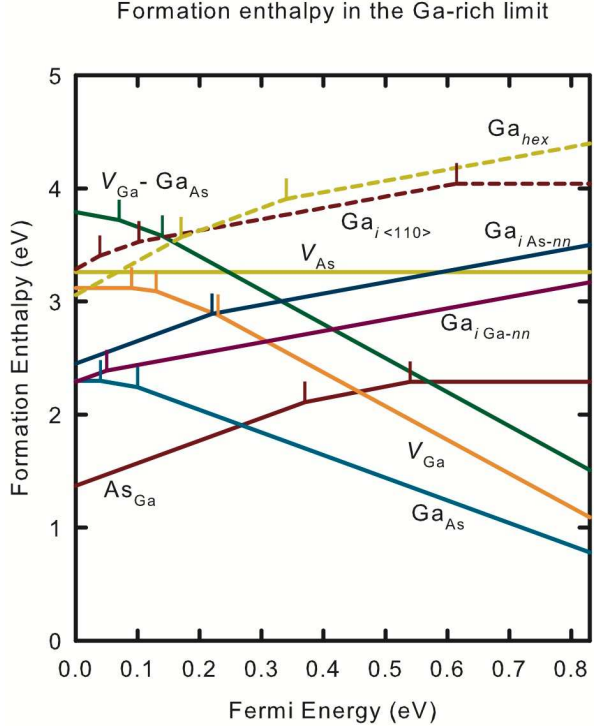}%
\caption{\label{figure:enthalpy}(Color online) 
Formation energies of low-energy point defects
and several gallium interstitials plotted as a function of Fermi energy across the
calculated band gap in the gallium-rich limit.  The meanings of the symbols are
$\mathrm{As_{Ga}} = $ arsenic antisite,
$\mathrm{Ga_{As}} = $ gallium antisite,
$V_\mathrm{Ga} = $ gallium vacancy,
$V_\mathrm{As} = $ arsenic vacancy,
$\mathrm{Ga}_{i\,\mathrm{As}-nn} = $ tetrahedral gallium interstitial with arsenic
nearest neighbors,
$\mathrm{Ga}_{i\,\mathrm{Ga}-nn} = $ tetrahedral gallium interstitial with gallium
nearest neighbors,
$\mathrm{Ga}_{i\,\langle 110 \rangle } = \langle 110 \rangle$ gallium-gallium split interstitial, 
$\mathrm{Ga}_\mathit{hex} = $ gallium interstitial in the hexagonal position,
and
$V_\mathrm{Ga}-\mathrm{Ga_{As}} = $ gallium vacancy created when an
arsenic vacancy is eliminated by being occupied with a neighboring gallium atom.
The $V_{\mathrm{Ga}}-\mathrm{Ga_{As}} $ defect was first identified as the lower-energy
configuration of the arsenic vacancy for Fermi energies higher in gap by
Baraff and {Schl\"uter}.\cite{Baraff1985b,Baraff1986}
Unstable barrier configurations are indicated with dashed lines.}
\end{figure}

\begin{figure}
\includegraphics{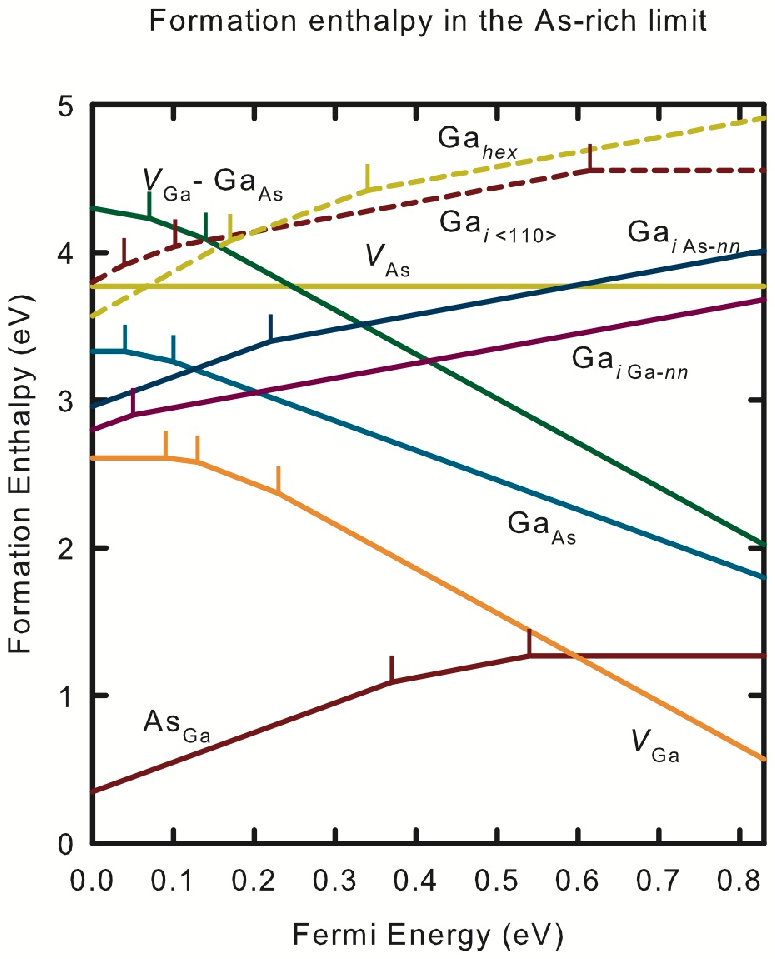}%
\caption{\label{figure:asrichenthalpy}(Color online)
Formation energies of low-energy point
defects and several gallium interstitials plotted as a function of Fermi energy across
the calculated band gap in the arsenic-rich limit.}
\end{figure}

Gallium diffusion rates can vary strongly due to varying concentrations
of the gallium defects which are available to diffuse.\cite{Tan1991a}
Gallium vacancy migration involves the motion of gallium atoms from a
nearby lattice site to the vacancy site.  The 
gallium atom moves in the opposite direction to that of the vacancy and
the gallium remains on the gallium sublattice.
Gallium interstitials are gallium atoms outside the crystal lattice and may
move from one position to another with or without entering the gallium sublattice.
Gallium antisite defects move either by the antisite atom moving into a nearby vacant
site or an interstitial location.  In order for gallium to move through the lattice as a gallium antisite, 
the antisite gallium atom must move into a nearby arsenic vacancy. 
However, as we will show below,
arsenic vacancies are not among the most numerous defects for any equilibrium conditions.  
Therefore adding the requirement of a nearby arsenic vacancy reduces
the ability for the gallium antisite to be an agent of gallium diffusion.  On the other hand,
if the antisite breaks apart into an independent arsenic vacancy and an
independent gallium interstitial, then gallium diffusion would proceed via an
interstitial pathway as described above.  So we focus on the basic processes of gallium interstitial 
diffusion in our
calculations and consider only the competition between gallium interstitial diffusion and 
gallium vacancy diffusion in our final analysis of
gallium diffusion.

Relative concentrations of gallium and arsenic in the environment when gallium arsenide 
is grown will affect the concentrations of defects containing an 
excess of gallium or arsenic which are 
present in the material after growth.
In other words, the equilibrium concentrations of defects involving an excess of
one of the native species or the other will
be controlled by the chemical potential.  Gallium arsenide 
material is stable when the chemical potential falls within the allowed 
compositional range, between the arsenic-rich and gallium-rich limits.  
The Fermi energy will also affect the equilibrium concentrations of most defects
because they can exist in many different charged states, 
which can be formed by donating or accepting electrons
from the crystal.  
Under equilibrium conditions we can estimate the expected
concentrations of each of these defects from their formation energies.

%

We present enthalpies of formation for
the complete set of point defects which may play a role in gallium diffusion, 
according to our calculations, as a function of Fermi energy in
Figs.\ \ref{figure:enthalpy} and \ref{figure:asrichenthalpy},
in the gallium-rich and arsenic-rich limits, respectively.  
In order to determine which defect configurations and charge states are likely to play an 
important role in gallium diffusion, we relaxed numerous initial defect configurations into their
lowest-energy geometries, using the approach and 
convergence limits described in
Section \ref{section:approach}.
\footnote{Differences in the formation energies presented
here and our previously published calculations\cite{Schick2002a}
are a result of having now evaluated the bulk arsenic structure
within our own computations, rather than relying upon a previous
calculation that used the same codes.  This brings our formation
energies into closer agreement with other work.  For example,
the neutral arsenic antisite formation energy in the arsenic-rich limit was stated to
be 1.8~eV in our earlier work, and with the updated bulk arsenic formation energy it is
now 1.3~eV, which agrees with the calculation due to Schultz {\em et al.}\cite{Schultz2009a}
to within the expected precision of density functional theory.}  
We have included in Figs.\ \ref{figure:enthalpy} and \ref{figure:asrichenthalpy} 
various configurations of gallium interstitials as 
well as gallium and arsenic vacancies and gallium and arsenic antisites, but we have 
not included arsenic interstitials, since they are not expected to play 
a direct role in gallium diffusion, and arsenic interstitial formation energies are 
large enough so that charged arsenic interstitial 
concentrations are too low to change the effective doping 
significantly enough to affect the concentrations of other charged defects over the entire range of 
conditions considered in Section \ref{subsection:equilibriumdefectconcentrations}, 
from $p$-type to $n$-type and from Ga-rich to As-rich.

The dependence of the formation energy for each defect upon the chemical
potential can be seen by comparing the energies of formation in
Figs.\ \ref{figure:enthalpy} and \ref{figure:asrichenthalpy} for the same defect
to each other.
Defects with excess gallium possess lower energies in Fig.\ \ref{figure:enthalpy},
which represents the gallium-rich limit, and higher energies in 
Fig.\ \ref{figure:asrichenthalpy}.  The opposite is the case for defects with an excess
of arsenic.

The formation energy of each specific defect in Figs.\ \ref{figure:enthalpy}
and \ref{figure:asrichenthalpy} is presented as
a continuous graph of a set of straight segments
representing the energy to form the
lowest-energy charge state of the
specified defect.  The slope of
each segment corresponds to the charge of the lowest-energy
state of the defect in units of the fundamental charge $e$.  The Fermi energies
corresponding to the points at which two of the straight segments
in the graph are joined are the charge transition levels.  The charge transition levels
are indicated with vertical lines in these figures.

\begin{figure}
\includegraphics{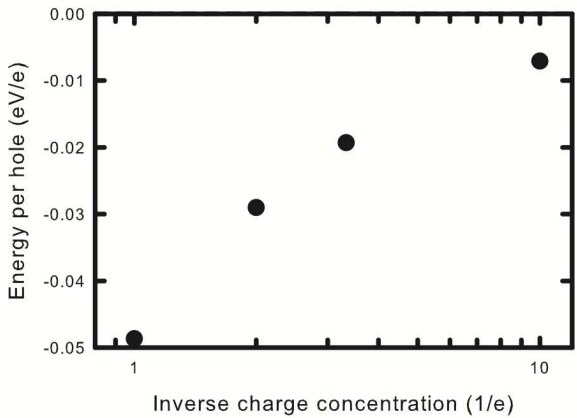}%
\caption{\label{figure:vbconverge}
The energy per hole for the transition between the neutral 216 atom GaAs supercell
and an equivalent supercell in a positive charge state $q$ is displayed as a function
of the reciprocal of the charge state.  The origin
of the energy axis is placed at the valence band edge, defined as the energy of the highest 
occupied Kohn-Sham state.
}
\end{figure}

The Fermi energy in Figs.\ \ref{figure:enthalpy} and \ref{figure:asrichenthalpy}
covers the range from 0~eV to 0.83~eV, corresponding to the energy range
from the highest occupied to the the lowest
unoccupied Kohn-Sham state (at the $\Gamma$ point) 
in the bulk calculation.  Lany and Zunger\cite{Lany2008a}  have justified the 
identification of this energy range as the calculated band gap 
by showing that valence band edge defined as the difference in 
energy between the neutral and 
the +1 state for the bulk semiconductor in the dilute limit, corresponding to adding one hole to 
an infinitely large bulk crystal, and the conduction band edge  
defined as the difference in energy between the $-1$ and the 
neutral state for the bulk semiconductor in the dilute limit, corresponding to adding one 
conduction electron to 
an infinitely large bulk crystal, are equal to the energies of the 
highest occupied and lowest unoccupied 
Kohn-Sham states at the $\Gamma$ point 
in the bulk calculation, respectively, in the direct-gap semiconductor ZnO.  We have verified that
the difference in 
energy between the neutral and 
the +1 state for the bulk semiconductor approaches the 
Kohn-Sham energy of the highest occupied state 
in the dilute limit and the difference in 
energy between the $-1$ and 
the neutral state for the bulk semiconductor approaches the 
Kohn-Sham energy of the lowest unoccupied state 
in the dilute limit in our calculations as well.  The results of this calculation for the 
valence band edge are shown in Fig.\ \ref{figure:vbconverge}.

At the current time, there is ongoing discussion about the underestimation of the band gap 
in standard density functional calculations and the effect of this on comparing density functional
results for defect properties to experimental 
results,\cite{Schultz2009a,Schultz2006a,Tuttle2008a,Schultz2008a} including whether and when 
a shift of transition levels by the so-called `scissors
operator'\cite{Baraff1984,Johnson1998a} is justified.
A full evaluation of every possible approach to correcting this problem is beyond the focus of
this paper.  Instead, we present the formation energies as they are calculated
and describe the character of the associated electronic states for those
cases in which there may be a special need to make corrections.
When performing the analysis for diffusion, we employ the as-calculated
numbers and also make reasonable estimates for the effects of corrections
for the few cases in which such corrections may be needed.  
We obtain clear predictions for diffusion from this set of calculations, 
which are in reasonable agreement with all other recent calculations of 
one or another of the quantities reported here.

\begin{figure}
\includegraphics{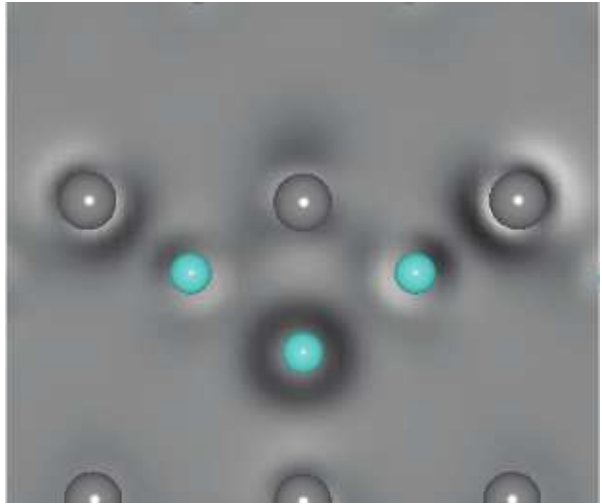}%
\caption{
\label{figure:chdone2two}
(Color online)
The difference in the distribution of electrons, when the tetrahedral gallium defect
charge state is changed from doubly positive to singly positive, is displayed in this
gray scale image.  The interstitial is shown in a (110) plane containing one of the
bonding chains of gallium and arsenic just above the interstitial.  The darker regions
indicate an increase in the electron density due to the transition from the more positive charge
state to the less positive charge state.  The lighter regions indicate a depletion of
electron density due to the same transition.  The very dark region surrounding the interstitial
atom indicates a substantial increase in electron density in this region due to the added electron.
}
\end{figure}
\begin{figure}
\includegraphics{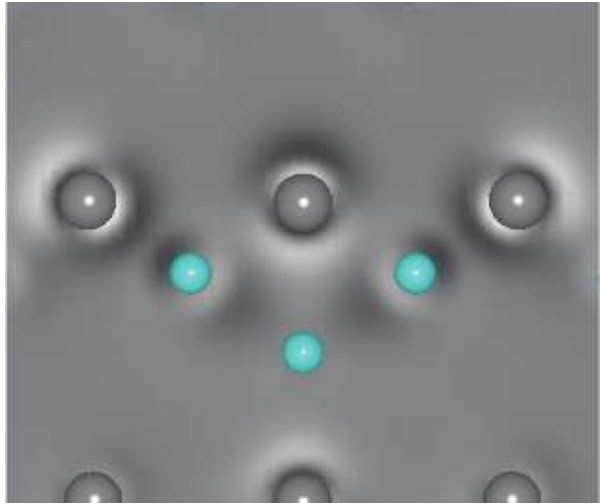}%
\caption{
\label{figure:chdneutral2one}
(Color online)
The difference in the distribution of electrons, when
the tetrahedral gallium defect charge state is changed
from singly positive to neutral, is displayed in this gray scale image.  
Here we see no large increase in the electron density in the immediate neighborhood of the interstitial due to the added electron.  Instead,
the added electron density is spread out across a wide region of the surrounding
bonding chains.
}
\end{figure}

Gallium interstitials are found to favor tetrahedral interstitial positions in our own calculations,
as in the results of Malouin \emph{et al}.\cite{Malouin2007a}
and Schultz \emph{et al.}\cite{Schultz2009a}
In our results, the gallium in the tetrahedral interstitial position surrounded
by gallium atoms possesses the lowest formation enthalpy for charge states from neutral through
doubly positive for all interstitial configurations. 
In the triply positive charge state, the tetrahedral positions between either
gallium or arsenic atoms have nearly the same energy cost.
Comparison of our formation energies to those of Schultz {\em et al.}, who finds that the 
tetrahedral gallium interstitial surrounded by arsenic atoms is the favorable
configuration for the non-zero charge states,\cite{Schultz2009a} 
shows differences of up to about 0.2~eV, an amount comparable to the
uncertainty of our density functional results, for all the point defects we can
compare.

The outer three electrons of the neutral gallium atom, which are involved in the bonding 
in GaAs, occupy 3$s$ (two electrons) and 3$p$ (the third electron) states in the case 
of an isolated gallium atom.  When this atom occupies a 
tetrahedral interstitial position in GaAs, 
it is far enough from the bonding chains that the two electrons filling the 3$s$ states, 
which lie closer to the gallium nucleus, may not be greatly perturbed by the surrounding 
lattice, while the last electron, which must occupy a higher energy state, may be 
strongly affected.

In order to examine the spatial distribution of the states occupied by these three outer 
electrons for a tetrahedral gallium interstitial in GaAs, we show the
redistribution of charge as the tetrahedral gallium interstitial between
gallium atoms makes charge state
transitions from doubly positive to singly positive and from singly positive to neutral,
respectively, in Figs.\ \ref{figure:chdone2two} and \ref{figure:chdneutral2one}.  
The dark regions in the figure represent regions in which negative
charge is added in each transition to the less positive charge state.
We have also examined the redistribution for the transition from triply positive
to doubly positive and found that this produced an image similar to
Fig.\ \ref{figure:chdone2two}, which we do not present here because of this
similarity.  

To understand these images it is easiest to begin with the triply positive
state, which is a gallium nucleus and its core electrons occupying the tetrahedral
interstitial position between gallium atoms.
The charge density changes
for each transition provide a map of the spatial distribution of the state occupied by
the added electron.  The first two electrons added, which make the transitions from
triply positive to doubly positive and then from doubly positive to singly positive,
can be seen to occupy $s$-orbital-like states, strongly concentrated in a 
spherically symmetric region close to the tetrahedral
gallium atom.  The image in Fig.\ \ref{figure:chdone2two} shows the electron going
into a tight, spherically symmetric region around the tetrahedral 
gallium atom, and not forming bonds with
neighboring atoms.  We conclude
that the associated transition levels correspond to strongly localized, low-energy $s$-like 
deep levels, and are correctly 
positioned with
respect to the valence band edge.

In order to examine the spatial distribution of the state occupied by the 
last of the three outer 
electrons, we show the
redistribution of charge as the tetrahedral gallium interstitial between
gallium atoms makes the 
transition from singly positive to neutral in Fig.\ \ref{figure:chdneutral2one}.  
We can see that 
this last electron occupies a wide region surrounding the defect, involving the 
neighboring atoms for some distance.   We conclude that 
any strongly localized, 3$p$-derived state for this last electron 
must be at a higher energy than the conduction band edge states in this calculation, and that the 
calculated transition level between the singly positive and neutral charge states 
for this interstitial, which is very near the calculated conduction band edge, 
actually corresponds to putting the final electron into conduction band edge states 
rather than into a localized deep level on the interstitial. This result is in agreement with 
the analysis presented in  Ref.\ \onlinecite{Schultz2009a}, which also concludes that 
the localized deep level corresponding to the neutral tetrahedral gallium interstitial between
gallium atoms must be a resonance in the conduction band. 
 
An electron at the conduction band edge in the vicinity of a singly positive 
tetrahedral interstitial can lower its energy by occupying a shallow hydrogenic bound state 
composed of conduction band edge states, so that there will be a transition to the neutral 
interstitial charge state just below the conduction band edge, with the final electron 
occupying such a shallow hydrogenic bound state.  Our estimates for such
a state using the experimental static dieletric constant of 
12.85 for GaAs\cite{Blakemore1982a}
give a binding energy of about 5.5\,meV with an effective Bohr radius
of around 100\,\AA, which is much larger than the dimension of the supercell used
in our calculations.  Since our density functional calculations put the last electron 
in the lowest energy state, this electron should be observed to occupy 
the shallow hydrogenic bound state.  However, due to size restrictions, the 
density functional calculation cannot properly represent the spatial 
distribution of this state.

Because the calculated energies of both the conduction band 
edge states and the hydrogenic bound state composed of these states are 
lower than they should be 
as a result of the known issues of density
functional theory in the local density approximation, the formation energy
of the neutral tetrahedral gallium interstitial between gallium atoms will be
underestimated.  The calculated formation energy for the neutral tetrahedral 
gallium interstitial between gallium atoms (with the last electron occupying a shallow 
hydrogenic bound state composed of conduction band states) should be shifted up 
by a scissors-type correction --- as is done for the conduction band edge states, to shift the 
calculated conduction band edge up to match the experimental gap --- in order to 
correctly represent a transition level just below the conduction band edge for the 
transition from singly positive to neutral, with the last electron occupying 
the shallow hydrogenic bound state.   

In order to calculate the scissors correction which must be applied to the 
formation energy of the neutral tetrahedral interstitial surrounded by gallium atoms, it is necessary 
to take into account the fact that the 
$\Gamma$ point is not included in the Monkhorst-Pack mesh\cite{Monkhorst1976}
we use for the summation over $k$-points 
in our defect calculations.  Therefore the energy of the lowest available conduction band states, 
out of which the hydrogenic bound state of the defect is formed, 
is higher than the calculated conduction band 
edge at the gamma point by 0.356 eV.  We determined this difference between the 
effective conduction band edge in our defect calculations and the calculated band edge at 
the $\Gamma$ point by taking the difference between the energy of the lowest conduction band 
states in the bulk calculation including the same $k$-points that we included in our 
defect calculations and the energy of the conduction band edge at the $\Gamma$ point from 
a bulk calculation.  The calculated +1/0 transition 
level for the neutral tetrahedral interstitial surrounded by gallium atoms is just below the 
effective conduction band edge for our defect calculations, at 0.330 eV 
above the calculated band edge at the $\Gamma$ point.   
Our calculations therefore give a binding energy of
26 meV for the electron bound to the neutral tetrahedral interstitial in the 
shallow hydrogenic bound 
state.  We do not expect our calculations to give a binding energy 
which matches the estimated binding energy 
of 5.5 meV for an electron bound to an isolated positively charged 
defect in a shallow hydrogenic state with an 
effective Bohr radius of around 100\,\AA, since the supercell dimensions in our defect calculations 
are much smaller than 100\,\AA, forcing the bound electron to be much 
closer than 100\,\AA to the nearest 
defect in the periodic array of defects in our supercells.

The scissors correction for the formation 
energy of the neutral tetrahedral interstitial surrounded by gallium 
atoms should be large enough to raise the 
+1/0 transition level from just below the calculated effective 
conduction band edge in our defect calculations to 
just below the experimental conduction band edge.  Since the experimental band gap depends 
on temperature, the scissors correction will also depend on temperature.  In our calculations of 
activation energies for diffusion, we will shift the formation energy of any 
neutral defect with its last electron occupying a hydrogenic 
bound state composed of the lowest available conduction band 
states up by the same scissors correction 
as is needed to shift the effective conduction band edge in our 
defect calculations up to coincide with the 
experimental conduction band edge.  Further corrections to remove the errors in the calculated 
binding energy of the hydrogenic state which occur due to the size of the supercells are not made, 
since they are within the estimated uncertainty of these calculations due to other causes.

As we have argued above, the existence of a shallow hydrogenic 
bound state for the gallium tetrahedral interstitial
between gallium atoms means that there must be a transition level between the singly positive 
state and the neutral state just below the conduction band edge. 
We also find a transition level between the triply- and doubly-positive charge states 
essentially at the valence band edge.   We conclude that the lowest energy interstitial
defect for an excess gallium atom in gallium arsenide is the tetrahedral
interstitial with four gallium nearest neighbors in the neutral,
singly-, doubly-, and triply-positive charge states,
depending on Fermi energy, across the calculated band gap.  As the Fermi energy 
approaches the valence band edge, the triply-positive and doubly-positive charge states 
become very nearly equally energetically favorable.  We also note that the formation energy 
for the tetrahedral gallium interstitial
surrounded by arsenic atoms in the triply-positive charge state is nearly the same
as that for the tetrahedral gallium interstitial surrounded by gallium atoms in this
charge state.

Just as the effective conduction band edge in our defect calculations is higher 
than the calculated conduction band edge at the $\Gamma$ point, the 
effective valence band edge in our defect calculations is lower than the calculated valence band 
edge at the $\Gamma$ point, due to
the fact that the Monkhorst-Pack mesh we use does not include the $\Gamma$ point.  
This helps to ensure that when the last two electrons are removed from the tetrahedral 
interstitial surrounded by gallium atoms, in order for this interstitial to 
make the transitions to the +2 and +3 charge 
states, these electrons are removed from highly localized 
defect states rather than delocalized valence band states.  
We can also see that these last two electrons 
are removed from localized defect states, producing defects which are truly in the +2 and +3 
charge states, by reviewing Fig.\ \ref{figure:chdone2two}, which shows the redistribution in charge 
resulting from the +2/+1 transition, and which looks essentially 
the same as a similar figure showing 
the redistribution in charge 
resulting from the +3/+2 transition.  We have not shown the figure for the +3/+2 transition 
due to this similarity.  Although the +3 charge state is not the charge state with the 
highest concentration for any Fermi level in the gap (since the +3/+2 
transition for the gallium tetrahedral interstitial
between gallium atoms occurs essentially at the valence band edge), we 
note that this charge state 
may still dominate diffusion if the migration barriers for diffusion in the +3 charge state are 
low enough that the total activation energy for diffusion (formation energy plus migration barrier) 
is lowest for these charge states.  For this reason, we consider diffusion in all charge 
states from +3 to neutral for the low-energy diffusion paths considered in this paper.

When we examine the properties of the tetrahedral gallium 
interstitial surrounded by arsenic atoms, 
we find that they are quite similar to the properties of the slightly more energetically favorable 
tetrahedral gallium interstitial surrounded by gallium atoms.
In particular, we find that the 
redistribution of charge as the tetrahedral interstitial between
arsenic atoms makes a 
transition from the doubly positive to the singly positive charge state 
is similar to the redistribution of 
charge as the tetrahedral interstitial between
gallium atoms makes this charge state
transition, as shown in Fig.\ \ref{figure:chdone2two} and discussed above.  This shows  
that the electron which is added in order to accomplish this charge transition goes into 
a highly localized state in the vicinity of the defect for both tetrahedral interstitials.  
We also find that the 
redistribution of charge as the tetrahedral interstitial between
arsenic atoms makes a 
transition from the singly positive to the neutral charge state is 
similar to the redistribution of charge as the tetrahedral interstitial between
gallium atoms makes this charge state
transition, as shown 
in Fig.\  \ref{figure:chdneutral2one} and discussed above.  This shows  
that the last electron which is added in order to accomplish the transition 
to the neutral charge state goes into 
a delocalized conduction-band-like state for both tetrahedral interstitials.  We conclude that 
the last electron goes into delocalized states at the conduction band edge in the supercell 
calculations for both of the neutral tetrahedral gallium interstitials, 
in agreement with Schultz {\em et al.}\cite{Schultz2009a}. 
 
As we have discussed above, in the vicinity of a  
defect in the +1 charge state, an electron at the conduction band edge can lower its energy 
by occupying a shallow hydrogenic bound state 
composed of conduction band edge states.  We conclude 
that there will be a transition to the neutral 
interstitial charge state just below the conduction band edge, with the final electron 
occupying such a shallow hydrogenic bound state, for the tetrahedral interstitial 
surrounded by arsenic atoms, as there is for the tetrahedral interstitial 
surrounded by gallium atoms.  As discussed above, the 
calculated formation energy for any neutral defect with its last electron occupying a shallow 
hydrogenic bound state composed of conduction band states should be shifted up 
by a scissors-type correction --- as is done for the conduction band edge states, to shift the 
calculated conduction band edge up to match the experimental gap --- in order to 
correctly represent a transition level just below the conduction band edge for the 
transition from singly positive to neutral.  As we do for the neutral tetrahedral interstitial 
surrounded by gallium atoms, in our calculations of 
activation energies for diffusion, we will shift the formation energy 
of the neutral tetrahedral interstitial surrounded by arsenic atoms 
up by the same scissors correction 
as is needed to shift the effective conduction band edge in 
our defect calculations up to coincide with the 
experimental conduction band edge.   

We find that the gallium hexagonal interstitial
configuration is unstable in all charge states across the 
band gap, and will relax to lower energy configurations if allowed, as first shown by
Malouin \emph{et al.}\cite{Malouin2007a}  By investigating the 
changes in the distribution of charge 
when the hexagonal gallium interstitial charge state is changed from 
singly positive to neutral, we find that the electron which is added to complete 
this transition occupies a delocalized conduction-band-like state, as was the case for 
the neutral tetrahedral interstitial surrounded by gallium atoms, discussed above.  
As in the earlier case, an electron at the conduction band edge in the vicinity of a singly positive 
hexagonal interstitial can lower its energy by occupying a shallow hydrogenic bound state 
composed of conduction band edge states, so that there will be a transition to the neutral 
interstitial charge state just below the conduction band edge, with the final electron 
occupying a shallow hydrogenic bound state. 
The calculated +1/0 transition 
level for the neutral hexagonal interstitial is just below the 
effective conduction band edge for our defect calculations, at 0.326 eV 
above the calculated band edge at the $\Gamma$ point, which gives a binding energy of
30 meV for the electron bound to the neutral hexagonal interstitial.
As in the case of the neutral tetrahedral interstitials, 
in our calculations of 
activation energies for diffusion, we will shift the formation energy of the 
neutral hexagonal interstitial up by the same scissors correction 
as is needed to shift the effective conduction band edge in 
our defect calculations up to coincide with the 
experimental conduction band edge.  

We find that the gallium bond-centered interstitial configuration
is also unstable in all charge states across the 
band gap, and will relax to lower energy configurations if allowed, as first shown by
Malouin \emph{et al.}\cite{Malouin2007a}.  When we relaxed the bond-centered 
interstitial, we discovered that
it generally relaxes in the direction of the tetrahedral interstitial, a view also
supported by the work of Malouin \emph{et al.}\cite{Malouin2007a}
In a slightly different result, Schultz \emph{et al.}\cite{Schultz2009a} found that 
the bond-centered gallium interstitial
relaxes to two configurations of lower symmetry which have been designated 
as gallium-arsenic split interstitials by Malouin \emph{et al.}\cite{Malouin2007a}  
In one of our calculations, we also observed a neutral 
bond-centered interstitial relaxing into a
gallium-arsenic split interstitial, in which the 
excess gallium atom shares an arsenic lattice site
with an arsenic atom. From these results, one may conclude that 
these split interstitial configurations are lower in energy than the original 
bond-centered configuration. 

A gallium split interstitial geometry involves an excess gallium atom
sharing a lattice site with the atom associated with that lattice
site.  The arrangement of the split interstitial atoms may be
symmetric with respect to exchange of the atoms if they are of
the same type, but the center of mass of the pair of atoms is
typically moved away from the lattice location.  Different gallium split interstitial configurations 
on same lattice site, having different orientations of the axis of the split 
interstitial pair and therefore different bonding with the atoms on the 
neighboring lattice sites, may have quite different energies.  

Metastable gallium split interstitial 
configurations which have been found in previous work\cite{Malouin2007a,Schultz2009a} 
include gallium-arsenic $\langle 110 \rangle$ and 
$\langle 111 \rangle$ split interstitials (using the nomenclature of 
Malouin \emph{et al.}\cite{Malouin2007a}), where the gallium atom shares an arsenic lattice site with the 
arsenic atom originally associated with that site, and gallium-gallium $\langle 110 \rangle$ and 
$\langle 100 \rangle$ split interstitials, where the gallium atom shares a gallium lattice site with the 
gallium atom originally associated with that site.  The same split interstitial configuration 
is not always found to be metastable in all the same charge states in 
References \onlinecite{Malouin2007a} and \onlinecite{Schultz2009a}.  
For example, the gallium-gallium $\langle 110 \rangle$ 
split interstitial is not identified as a metastable configuration for any charge state and is therefore not studied 
by Malouin \emph{et al.}\cite{Malouin2007a}. 

However, both of these earlier papers (and our work) 
agree on the general ordering of these split interstitial configurations, where they have been calculated.  
For example, the gallium-gallium 
$\langle 100 \rangle$ split interstitial has been found to be the highest in energy (or at least tied for the highest) 
in the neutral and positive charge states.\cite{Malouin2007a,Schultz2009a}  Both 
Malouin \emph{et al.}\cite{Malouin2007a} and Schultz \emph{et al.}\cite{Schultz2009a} identify the configuration 
called a $\langle 111 \rangle$ gallium-arsenic split interstitial by Malouin \emph{et al.}\cite{Malouin2007a} 
as the 
lowest energy non-tetrahedral configuration for the +1 charge state, with a formation energy 0.93 eV above 
the formation energy of the most energetically favorable tetrahedral interstitial according to 
Malouin \emph{et al.},\cite{Malouin2007a} and 0.95(0.83)eV above 
the formation energy of the most energetically favorable tetrahedral interstitial according to the 
LDA(PBE) calculations of 
Schultz \emph{et al}.\cite{Schultz2009a}  For the +2 and +3 charge states, Malouin \emph{et al.}\cite{Malouin2007a} 
identify the $\langle 111 \rangle$ gallium-arsenic split interstitial as the lowest energy 
non-tetrahedral configuration, at 0.93 eV above 
the formation energy of the most energetically favorable tetrahedral interstitial (for both charge states), 
but they have no calculations for the gallium-gallium $\langle 110 \rangle$ split interstitial.  
Schultz \emph{et al.}\cite{Schultz2009a} 
identify the gallium-gallium $\langle 110 \rangle$ split interstitial as the lowest energy 
non-tetrahedral configuration for the +2 and +3 charge states, at 0.9-1.0 eV higher than  
the formation energy of the tetrahedral interstitial surrounded by arsenic atoms, but they do not 
report any calculations for gallium-arsenic interstitials in these charge states, since they 
do not find the gallium-arsenic split interstitial configurations to be metastable for these charge states.

Although Malouin \emph{et al.}\cite{Malouin2007a} do not consider the gallium-gallium 
$\langle 110 \rangle$ split interstitial 
configuration, 
both Schultz \emph{et al.}\cite{Schultz2009a} and our present work find this configuration to be the 
lowest energy metastable non-tetrahedral configuration for the neutral charge state.  We find the 
formation energy of the neutral 
$\langle 110 \rangle$ split interstitial to be only 0.53(0.57) eV above 
the formation energy of the most energetically favorable tetrahedral interstitial, according to 
our norm-conserving pseudopotential (\textsc{vasp}) calculations.  Schultz \emph{et al.}\ report 
the formation energy of the neutral 
$\langle 110 \rangle$ split interstitial to be 0.6-0.7 eV above 
the formation energy of the most energetically favorable tetrahedral interstitial.\cite{Schultz2009a} 
The next higher energy configuration found for the neutral gallium interstitial is the 
$\langle 111 \rangle$ gallium-arsenic split interstitial, with a formation energy reported at 0.82 eV 
above the formation energy of the most energetically favorable tetrahedral interstitial 
by Malouin \emph{et al.},\cite{Malouin2007a} and 0.9 eV 
above the formation energy of the most energetically favorable tetrahedral interstitial 
by Schultz \emph{et al.}\cite{Schultz2009a}

We note that for all non-neutral charge states that seem 
likely to play an important role in gallium 
interstitial diffusion ($q$ = +1, +2, and +3), the formation energy 
of the lowest energy metastable non-tetrahedral configuration found by us or in previous work\cite{Malouin2007a,Schultz2009a} 
is 0.8-0.9 eV above the energy of the most 
energetically favorable tetrahedral interstitial.  We conclude that the energy barrier for 
neutral gallium interstitial diffusion via a path going through the gallium-gallium $\langle 110 \rangle$ 
split configuration may be lower than the energy barrier for any other charge state and diffusion pathway.  
We will therefore investigate the $\langle 110 \rangle$ split interstitial more closely.

\begin{figure}
\includegraphics{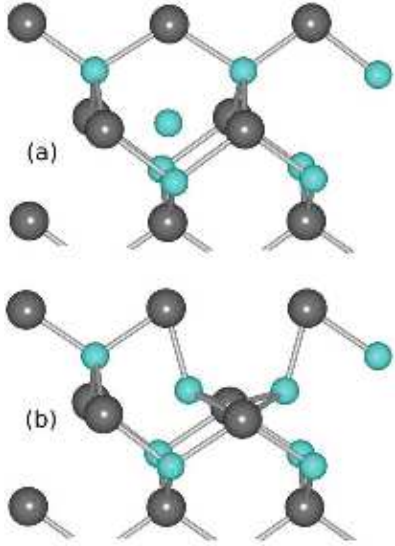}%
\caption{\label{figure:compareinterstitials}
(Color online) Comparison of neutral
gallium interstitials in the tetrahedral position between gallium atoms
(a) and in the gallium-gallium $\langle 110 \rangle$ 
split configuration (b).
The split interstitial is metastable in this charge state.
Gallium atoms are represented
with smaller light blue spheres and the arsenic atoms with larger dark gray spheres.
A small number of the lattice atoms surrounding the interstitial are presented in
these figures for the sake of clarity.}
\end{figure}

The neutral tetrahedral gallium interstitial surrounded by gallium atoms and the 
neutral gallium-gallium $\langle 110 \rangle$ split interstitial configurations are 
compared in Fig.\ \ref{figure:compareinterstitials}.  The neutral gallium
interstitial with gallium nearest neighbors is shown in
Fig.\ \ref{figure:compareinterstitials}(a), where
the interstitial is seen as an isolated atom to the left of the center, 
surrounded by four gallium atoms.
Additional atoms from the lattice are shown around the defect to enable ease of
comparison to the $\langle 110\rangle$ split interstitial that is shown in
Fig.\ \ref{figure:compareinterstitials}(b).

As can be seen in Fig.\ \ref{figure:compareinterstitials}(a), tetrahedral 
gallium interstitials maintain the symmetry of
the ideal tetrahedral defect, with an outward
relaxation of the neighboring atoms.  
There is an outward relaxation of the four nearest neighbor
gallium atoms by about 6.3\%, to a distance of 2.55\,{\AA}
from the neutral tetrahedral gallium interstitial.
(This may be compared to the 2.48\,{\AA} distance between the
nearest-neighbor atoms in bulk gallium.)
The bulk nearest neighbor distance in GaAs is 2.40\,{\AA}, as determined with the
same code and within the same standards of convergence described previously.
The distances between
the pairs of nearest-neighbor lattice gallium atoms around the defect is 4.17\,{\AA},
where the corresponding bulk next-nearest neighbor distance is
3.96\,{\AA}, {\em i.e.}\,there is a 5.3\% expansion.

For the neutral tetrahedral gallium interstitial with arsenic nearest neighbors,
the As neighbor atoms relax by about 5.4\% outward to a
distance of 2.53\,{\AA}.  The spacing between the four nearest-neighbor As atoms
surrounding the interstitial in this case is 4.14\,{\AA}, an expansion of 4.5\%.

\begin{figure}
\includegraphics{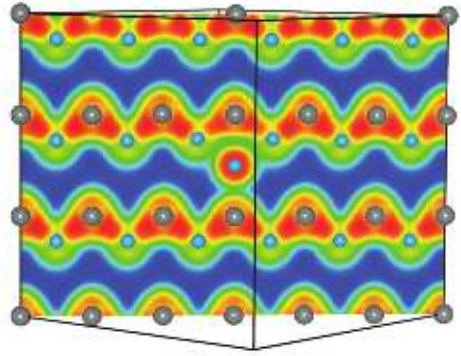}%
\caption{\label{figure:ELFtet}
(Color online)
The electron localization function\cite{Becke1990a} 
in a supercell containing a neutral gallium tetrahedral interstitial
surrounded by gallium nearest
neighbors is displayed for a $\left( 110 \right)$ plane passing through the
interstitial atom.
Gallium atoms are represented with smaller blue spheres and arsenic atoms,
with larger gray spheres.  Atoms between the
plane and the viewer have been removed for the sake of clarity.  
Higher values of the electron localization
function are shown in red.  The color yellow corresponds to the value of 0.5, below
which the localization function is too weak to indicate bonding. 
The interstitial is seen between two parallel bonding chains
near the center of the figure.   
Two of the nearest-neighbor gallium atoms are visible immediately above the defect,
just to the left and right.  The remaining two nearest neighbor gallium 
atoms are associated with a chain perpendicular to the plane shown here, 
and belong to a bonding chain
that includes the arsenic atom immediately below the interstitial.  There is little
evidence of bonding between the tetrahedral interstitial and any of the nearby
atoms.}
\end{figure}

A picture of the bonding in the neighborhood of a defect can be obtained by plotting
the electron localization function,\cite{Becke1990a} which is shown in Fig.\ \ref{figure:ELFtet}
for the neutral tetrahedral interstitial with gallium nearest neighbors.  As can be seen in
Fig.\ \ref{figure:ELFtet}, the interstitial gallium atom is not
bonded to the four nearest neighbor gallium atoms.  This is consistent with a picture
in which the highest occupied state is a hydrogenic level that is spread over
the volume of the supercell.  We can also see a high amount of electron localization in a
small, spherically symmetric region surrounding the interstitial atom, indicating that the
other two valence electrons which are bound to this neutral interstitial are in highly
localized, spherically symmetric, $s$-like states.

\begin{figure}
\includegraphics{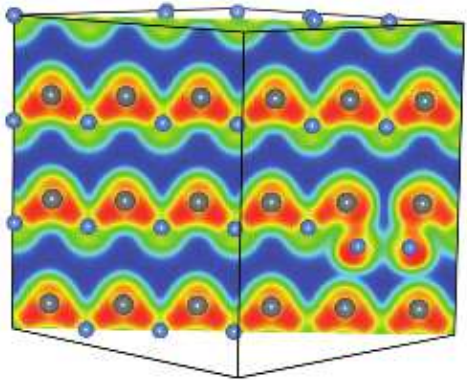}%
\caption{\label{figure:ELFinbox}
(Color online)
The electron localization function\cite{Becke1990a} 
in a supercell containing a neutral $\langle 110 \rangle$ Ga-Ga
split interstitial is displayed for a $\left( 110 \right)$ plane passing through the
interstitial atom.
Gallium atoms are represented with smaller blue spheres and arsenic atoms,
with larger gray spheres.  Atoms between the
plane and the viewer have been removed for the sake of clarity.  
Higher values of the electron localization
function are shown in red.  The color yellow corresponds to the value of 0.5, below
which the localization function is too weak to indicate bonding.   The defect is located in the lower
right quadrant of the figure where two gallium atoms are seen to be displaced
downward from the middle chain of bonds.  The two gallium atoms of the split interstitial are
displaced symmetrically from a gallium lattice site on this bonding chain.  
The results show that the two gallium atoms of the split 
interstitial are not bonded to each other, 
while each of these gallium atoms is strongly bonded to the 
nearest arsenic atom along the chain.}
\end{figure}

\begin{figure}
\includegraphics{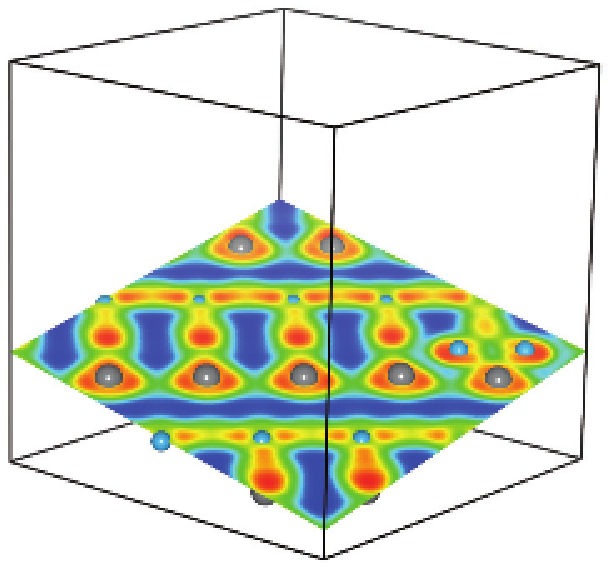}%
\caption{\label{figure:ELFtriangle}
(Color online)
The electron localization function\cite{Becke1990a} 
in a supercell containing a neutral $\langle 110 \rangle$ Ga-Ga
split interstitial is displayed for a $\left( 110 \right)$ plane 
that passes through the two gallium atoms of the
interstitial as well as one
of the two nearest-neighbor arsenic atoms in the bonding chain perpendicular 
to the axis of the split interstitial.  For clarity, only atoms very 
near this plane are shown.  Some bonding
is visible between each of the two gallium atoms of the defect
and the nearest neighbor arsenic atom of the perpendicular chain.}
\end{figure}

The metastable neutral gallium-gallium $\langle 110 \rangle$ 
split interstitial configuration is shown 
in Fig.\ \ref{figure:compareinterstitials}(b).
The center of the gallium split-interstitial pair is 0.85\,{\AA} below the
ideal gallium lattice site, as shown in Fig.\ \ref{figure:compareinterstitials}(b).  The distance 
between the two gallium atoms of this split interstitial is 2.74\,{\AA}, which is 15\%
farther than the calculated bulk gallium arsenide nearest 
neighbor distance of 2.40\,{\AA}.  Either arsenic atom shown 
bonding to, and above, the split interstitial in Fig.\ \ref{figure:compareinterstitials}(b) is
2.46\,{\AA} from the nearest gallium atom of the split pair, and 
the two arsenic atoms shown bonding to, and below, the
defect are 2.58\,{\AA} from either gallium in the defect.  Each 
split-interstitial gallium atom is 2.83\,{\AA} 
from the nearest other gallium atom in the crystal.  
We note that the gallium-gallium $\langle 110 \rangle$ split interstitial is closer to the original
gallium lattice site, and the distance between the two atoms of the gallium split-pair 
is larger than the corresponding distances\cite{Schick2002a} 
in the arsenic-arsenic $\langle 110 \rangle$
split interstitial.
  
We may also compare the distance between the two 
gallium atoms of the split interstitial to interatomic 
distances in the room-temperature phase
($\alpha$-gallium) of pure gallium.  In $\alpha$-gallium, 
the atoms have one nearest neighbor at 2.39\,{\AA} and four 
next-nearest neighbors at distances ranging from 2.71\,{\AA} to 
2.80\,{\AA}.\cite{Wyckoff1963a}  We find that the gallium split-pair atoms are separated from each other
by a distance within the range of the next-nearest neighbors in bulk gallium.

The lattice is expanded slightly and neighboring atoms move outward around a split interstitial, 
as they do around a tetrahedral interstitial.  
The distance between the two arsenic atoms above 
the split interstitial in Fig.\ \ref{figure:compareinterstitials}(b)
is 4.02\,{\AA}.
The distance between the two arsenic atoms shown below the split interstitial
in Fig.\ \ref{figure:compareinterstitials}(b) is 4.18\,{\AA}.
The distance between any one of the arsenic atoms below the split interstitial in the figure
and either one of the two arsenic atoms above the interstitial is 4.15\,{\AA}.
(For comparison, the calculated bulk gallium arsenide 
next-nearest neighbor distance is 3.92\,{\AA}.)   

As we can see in Fig.\ \ref{figure:compareinterstitials}, the extra 
gallium atom does not have to move far for the 
tetrahedral interstitial between gallium atoms to transform into a 
gallium-gallium $\langle 110 \rangle$ 
split interstitial.  In moving from the tetrahedral interstitial
location to become part of the split interstitial defect, the Ga atom is displaced
a total of 0.81\,{\AA}.

The bonding between the two gallium atoms of the neutral $\langle 110 \rangle$ 
split interstitial, and between these atoms and their arsenic nearest neighbors, 
is investigated in Fig.\ \ref{figure:ELFinbox} and Fig.\ \ref{figure:ELFtriangle}.  
The electron localization function is shown for a plane cutting through
the split interstitial and the $\langle 110 \rangle$ chain parallel to the 
axis of the split interstitial in Fig.\ \ref{figure:ELFinbox}.  This figure shows that  
the two gallium atoms of the interstitial are strongly bonded 
to the nearest arsenic atoms in this chain, but they are not bonded to each other.  
The visible lack of bonding between the two gallium atoms of the 
split interstitial in this figure is not very surprising, since we have mentioned above that the 
distance between these two atoms is 
larger than the nearest-neighbor distance in GaAs, the 
distance between the two arsenic atoms of the equivalent 
arsenic-arsenic $\langle 110 \rangle$ split interstitial in GaAs, and the nearest neighbor distance 
in the room-temperature phase of bulk gallium. 

In addition to the strong bonding between the gallium atoms of the split interstitial and 
their arsenic nearest neighbors on the $\langle 110 \rangle$ chain parallel to the 
interstitial axis, shown in Fig.\ \ref{figure:ELFinbox}, there is some weaker bonding between 
both gallium atoms of the split interstitial
and the other two arsenic nearest neighbors of the interstitial lattice site, 
which are on the bonding chain perpendicular to the
axis of the split interstitial.  This bonding is visible in Fig.\ \ref{figure:ELFtriangle}, which shows 
the electron localization function in the plane that
cuts through the split interstitial and one of these arsenic atoms (the next arsenic atom closer
to the viewer).

\begin{figure}
\includegraphics{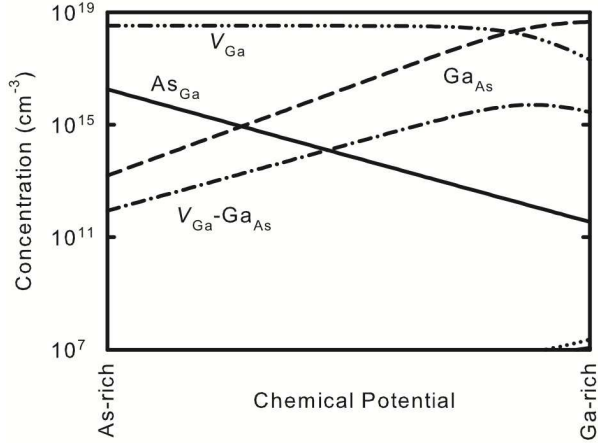}%
\caption{\label{figure:ntypeconcentrations}Calculated equilibrium 
concentrations of the most common point defects
at 1100 K in \emph{n}-type gallium arsenide with a net donor 
concentration $N_d = 10^{19}\ \mathrm{cm}^{-3}$, as a
function of chemical potential across the stoichiometric range.}
\end{figure}

\begin{figure}
\includegraphics{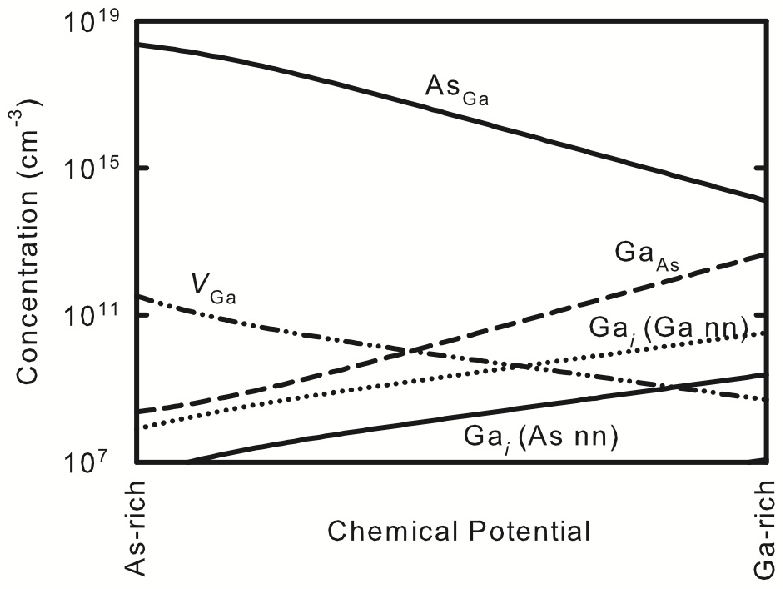}%
\caption{\label{figure:ptypeconcentrations}Calculated equilibrium
concentrations of the most common point defects
at 1100 K in \emph{p}-type gallium arsenide with a net donor 
concentration $N_d = -10^{19}\ \mathrm{cm}^{-3}$, as a function
of chemical potential across the stoichiometric range.}
\end{figure}

\begin{figure}
\includegraphics{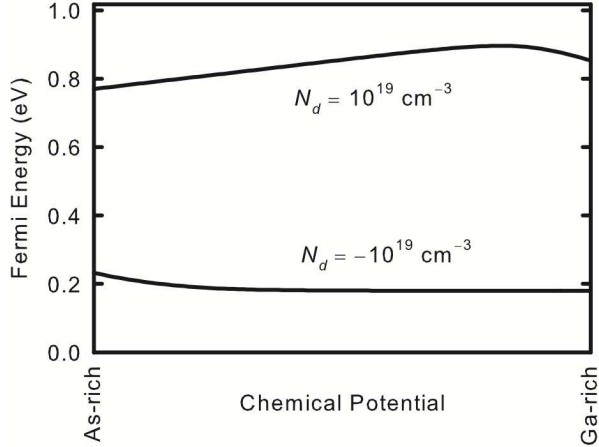}%
\caption{\label{figure:fermienergy}Fermi energies in \emph{p}- and \emph{n}-type
gallium arsenide as a function of chemical potential across the stoichiometric range.  
The full experimental band gap at 1100 K extrapolated from experiment 
is presented on the vertical axis.\cite{Blakemore1982a}
The lower energy curve is the Fermi energy for the \emph{p}-type material
and the higher energy curve is the Fermi energy for the \emph{n}-type material,
as indicated by the doping levels $N_d$.
Both are calculated at a temperature of 1100\,K.}
\end{figure}

\subsection{\label{subsection:equilibriumdefectconcentrations}Equilibrium defect concentrations}

The concentrations of defects in the material under 
conditions of thermal equilibrium at a given temperature are 
determined by a self-consistent procedure 
in which a full account is taken of the effective doping
concentration, chemical potential, and overall charge neutrality.
When the chemical
potential is within the range in which gallium arsenide can form, then
the majority of the atoms fall into the usual lattice for that material.
The concentration of a particular defect in a specific charge state is
proportional to a Boltzmann factor,
\begin{equation}
N_s e^{-G_{f,i}(N_\mathrm{Ga},N_\mathrm{As},N_\mathrm{e}) / kT}\, ,
\end{equation}
where $G_{f,i}(N_\mathrm{Ga},N_\mathrm{As},N_\mathrm{e})$
is the formation energy for defect $i$ in charge state $q=-N_\mathrm{e}$ 
calculated from Eq.\,\ref{eq:gibbs}, the net excess (or deficit) of 
Ga atoms involved in the defect is given by $N_\mathrm{Ga} - N_\mathrm{As}$,
and $N_s$ is the concentration of sites available for this defect.  
(Please note that the specification of the defect also includes its 
configuration: \emph{e.g.}\ a gallium interstitial in a tetrahedral site 
with gallium nearest neighbors.)  
This formation energy depends explicitly on the chemical potential and Fermi energy.

We determine the concentrations of all the defects by requiring 
charge neutrality as given in the following equation:
\begin{eqnarray}
\label{eq:selfconsistent}
N_d = \sum_{i,N_\mathrm{e}} N_\mathrm{e} N_s
e^{-G_{f,i}(N_\mathrm{Ga},N_\mathrm{As},N_\mathrm{e})/kT} \nonumber \\
 - N_v\,e^{-\epsilon_F/kT}
+ N_c\,e^{-(E_g-\epsilon_F)/kT}\, ,
\end{eqnarray}
where the effective doping concentration,
$N_d$, is given by the concentration of ionized donor impurities minus 
the concentration of ionized acceptor impurities.\cite{Zhang1991,Schick2002a}
The second and third terms on the right hand side of Eq.\,\ref{eq:selfconsistent} 
correspond to subtracting the density of holes in the valence band and adding the 
density of electrons in the conduction band, respectively, with $N_v$ and $N_c$ 
determined from experiment as described in 
Ref.\ \onlinecite{Blakemore1982a}.  

At each temperature, this calculation is performed for 
values of the chemical potential
over the entire compositional range from arsenic rich to gallium rich,
giving the value of the Fermi energy and the resulting concentrations of all the 
charged defects, which depend on the Fermi energy.
In doing this calculation, the band-gap in gallium arsenide is taken to be the
experimentally determined gap or the gap extrapolated from 
experiment,\cite{Blakemore1982a} and the defect formation
energies are those produced by the present density-functional calculation.
Scissors corrections for charge state transition levels involving movement of 
an electron to or from a localized state with conduction-band-like character, 
such as the donor levels of the arsenic antisite, do not make a qualitative difference in 
Figs.\ \ref{figure:ntypeconcentrations} to 
\ref{figure:fermienergy}; they do not change the ordering of the most numerous defects. 

Our estimated equilibrium concentrations are displayed
for an annealing temperature of 1100 K for both strong $n$-type 
and strong
$p$-type doping in Figs.\ \ref{figure:ntypeconcentrations} and 
\ref{figure:ptypeconcentrations}, respectively.
When the material is doped $n$-type, the defects that have acceptor levels
become energetically more favorable, yielding higher concentrations of these
defects relative to the intrinsic case.
Consistent with this expectation and the fact that the gallium antisite and
gallium vacancy possess acceptor levels,
we see in Fig.\ \ref{figure:ntypeconcentrations} that these are the
dominant defects in $n$-type gallium arsenide.
For strongly $p$-type material, we see in Fig.\  
\ref{figure:ptypeconcentrations} that 
the arsenic antisite, a donor, is the only dominant defect species.

At an annealing temperature of 1100\,K the Fermi energy
for the $n$-type material varies between 0.77~eV and 0.90~eV, while the
Fermi energy for the $p$-type material varies between 0.18~eV and 0.22~eV, as
shown in Fig.\ \ref{figure:fermienergy}.
Without the compensating effect of the defects, the Fermi energy would
be 0.18~eV for the $p$-type material and very near the conduction band edge for the 
$n$-type material.
In $n$-type material, the Fermi energy shows signs of strong compensation 
across the entire compositional range from gallium rich
to arsenic rich, 
with especially strong compensation closer to the arsenic-rich limit, 
due to large concentrations of gallium vacancies.  
In $p$-type material near the arsenic-rich limit, there is 
evidence of compensation due to the arsenic antisites.

In $n$-type material, the gallium vacancy is seen in
Fig.\ \ref{figure:ntypeconcentrations} to
be present in high concentrations, while
gallium interstitial concentrations are negligible, falling well below
the lower limit of the graph.
In $p$-type GaAs, as seen in Fig.\ \ref{figure:ptypeconcentrations}, 
the gallium vacancy occurs in far lower concentrations --- 
around two orders of magnitude lower than gallium interstitial concentrations
in the Ga-rich limit.  We conclude that Ga interstitials would be
likely to play an important role in the diffusion of Ga in $p$-type, Ga-rich GaAs.

Since equilibrium concentrations of gallium interstitials are still 
rather low even in the $p$-type material, we would expect
only moderate gallium diffusion rates due to interstitials.
Diffusion rates of gallium interstitials may be significantly enhanced
if gallium atoms are knocked out of their lattice by ion implantation
or irradiation, creating more significant concentrations of gallium
interstitials.
To quantitatively evaluate
the relative contribution of gallium interstitials to gallium diffusion
under different experimental conditions,
we must compute the barriers for motion between
stable configurations of gallium interstitials, which requires us to survey
numerous possible migration pathways.

\subsection{\label{subsection:pathways for diffusion}Pathways for diffusion}

Previously, migration barriers for two low energy migration 
paths for diffusion of gallium interstitials have
been calculated for the +1 and neutral charge states.\cite{Levasseur-Smith2008a}  
In both cases, an automated procedure was 
initially used to identify a low energy diffusion path, and the results for the 
saddle point energy were then converged further.  In one case, a low energy 
path was discovered for a gallium atom to move between the 
tetrahedral interstitial site between gallium atoms and the tetrahedral
interstitial site between arsenic atoms through the connecting hexagonal interstitial
site, without entering the lattice network.  And in the other case (for the +1 charge state only), 
a low energy path was discovered for the gallium 
atom to move from a gallium-arsenic 
$\langle 111 \rangle$ split interstitial to the tetrahedral interstitial
site surrounded by gallium atoms.\cite{Levasseur-Smith2008a} 
This path can be the second step in the migration of a gallium atom from one tetrahedral site 
surrounded by gallium atoms to 
another equivalent tetrahedral site.  The first step for this migration would require the 
gallium atom to reverse this motion, leaving  
the initial tetrahedral site to form  
a $\langle 111 \rangle$ gallium-arsenic split interstitial. 

The energy barriers calculated for these two pathways for diffusion in the 
neutral and +1 charge states were quite similar. 
The energy barriers calculated by Levasseur-Smith \emph{et al.}\cite{Levasseur-Smith2008a} for 
migration of a gallium atom from the tetrahedral site surrounded by gallium atoms 
through the hexagonal site to the tetrahedral site surrounded by arsenic 
atoms (and on through another hexagonal site to a different 
tetrahedral site surrounded by gallium atoms) are 1.3 eV for the neutral 
charge state and 1.4 eV for the +1 charge state.  The energy barrier calculated 
by Levasseur-Smith \emph{et al.}\cite{Levasseur-Smith2008a} for 
migration of a gallium atom from a tetrahedral site surrounded by gallium atoms 
through the gallium-arsenic 
$\langle 111 \rangle$ split interstitial configuration to another tetrahedral interstitial
site surrounded by gallium atoms is 1.4 eV (composed of an energy 
difference of 0.9 eV between the 
$\langle 111 \rangle$ split interstitial and the tetrahedral interstitial 
surrounded by gallium atoms and an additional barrier of 0.4 eV 
to move between these configurations) 
for the +1 charge state.
Levasseur-Smith \emph{et al.}\ only used their automated procedure 
to investigate the low energy paths for 
migration of a gallium atom out of the tetrahedral interstitial configurations (in the neutral and +1 
charge states) and the 
gallium-arsenic $\langle 111 \rangle$ split interstitial configuration (in the +1 charge state), 
since the other 
metastable gallium-gallium and gallium-arsenic split interstitial configurations 
discovered by this group\cite{Malouin2007a} for the neutral and 
positive charge states were at least 0.8 eV 
higher in energy than the tetrahedral interstitial configuration surrounded by 
gallium atoms.\cite{Levasseur-Smith2008a} 

Schultz \emph{et al.}\ identified other metastable 
gallium interstitial configurations and also calculated the 
energy barrier for gallium diffusion between tetrahedral interstitial configurations 
via the non-network path through 
the hexagonal saddle point in the +1 charge state, obtaining 
1.22 eV (1.11 eV) in LDA (PBE) calculations.
\cite{Schultz2009a} Based on these results for the migration barrier 
through the hexagonal site and 
the results of Levasseur-Smith \emph{et al.}\ for the migration barrier through the 
gallium-arsenic 
$\langle 111 \rangle$ split interstitial configuration, 
together with their calculated formation energies in many charge states 
for a large number of metastable defects 
involving insertion of the excess gallium atom into the lattice, Schultz \emph{et al.}\
predicted that the dominant path for gallium interstitial diffusion in the +1 charge state 
is likely to be along the non-network path through the hexagonal site, with 
smaller contributions from in-network paths, and that the migration barriers for the 2+ 
and 3+ states should be similar to the migration barriers for the 1+ states.\cite{Schultz2009a}
Given the extent of the survey of defect formation energies available to us,
we can now more fully investigate possible low-energy pathways for gallium interstitial
diffusion in all the charge states likely to be important for a Fermi level anywhere in the gap.  

According to the results presented in Section \ref{subsection:energyandstructure}, diffusion 
of the neutral gallium interstitial via 
a path through the gallium-gallium $\langle 110 \rangle$ split interstitial configuration 
may have a lower energy barrier than diffusion along other paths and for other charge states.  
As we have noted, the increase in energy above the most 
energetically favorable tetrahedral interstitial is lower for the 
neutral gallium-gallium $\langle 110 \rangle$ split interstitial than for any other metastable 
neutral or positively charged gallium split interstitial found in earlier 
work.\cite{Malouin2007a,Schultz2009a} 
Using results of our density functional calculations with norm-conserving pseudopotentials, we 
found that the formation energy of the metastable neutral gallium-gallium $\langle 110 \rangle$ 
split interstitial
is only 0.53~eV higher than
the formation energy calculated for the lowest energy neutral tetrahedral gallium interstitial 
--- this energy difference is 0.57~eV according to the \textsc{vasp} calculations 
which we used in conjunction 
with the nudged elastic band method to determine minimum 
energy diffusion paths and migration barriers, and it is 0.6-0.7 eV according to 
Schultz \emph{et al.}\cite{Schultz2009a}  Our discussion in the rest of the 
paper will be based on the 
calculated \textsc{vasp} results for 
the formation energies and migration barriers. 

The initial and middle steps for gallium diffusion between 
two tetrahedral interstitial sites surrounded by 
gallium atoms through the gallium-gallium $\langle 110 \rangle$ 
split interstitial configuration can be seen 
in Figs.\ \ref{figure:compareinterstitials}(a) and \ref{figure:compareinterstitials}(b), respectively.
The tetrahedral gallium interstitial atom and the nearest neighbor gallium atom above it
and to the right in Fig.\ \ref{figure:compareinterstitials}(a)
move a short distance to the right, with a corresponding upward or downward motion,
to form the $\langle 110 \rangle$ split pair as seen in
Fig.\ \ref{figure:compareinterstitials}(b).
The rest of the motion continues the general trend with the right
atom of the split interstitial falling into the nearest tetrahedral defect position
to the right, while the original gallium interstitial atom moves onto the
gallium lattice site beside the new tetrahedral interstitial. 
The net result is the motion of one gallium atom from the original tetrahedral 
interstitial site to an equivalent neighboring tetrahedral interstitial site.

We used the nudged elastic band 
method\cite{Jonsson1998a,Mills1995a,vasp1,vasp2,vasp3,vasp4}
to determine the energy barriers for the migration of a gallium atom starting in the 
tetrahedral gallium interstitial configuration with gallium nearest neighbors, through both the
gallium-gallium $\langle 110 \rangle$ split-interstitial pathway and the hexagonal
pathway, for the neutral, +1, +2, and +3 charge states.  Only the end points 
(tetrahedral interstitial configurations) were held
fixed, and the pathway was allowed to converge freely.
The calculated barriers range from 0.6~eV to 1.2~eV, as displayed in 
Table \ref{table:migrationenergies}.
The lowest barrier corresponds to the gallium interstitial passing through the
gallium-gallium $\langle 110 \rangle$ split interstitial position in the neutral charge state.  

\begin{table}[floatfix]
\caption{Uncorrected migration barriers for gallium interstitial diffusion
from one tetrahedral position to the nearest neighboring tetrahedral
interstitial position, as calculated using the nudged elastic band method.}
\label{table:migrationenergies}
\begin{ruledtabular}
\begin{tabular}{ccc}
Charge state&$\langle 110 \rangle$-split pathway&hexagonal pathway\\
 &barrier (eV)&barrier (eV)\\
\hline
$0$&$0.6$&$1.2$\\
$1+$&$1.1$&$1.2$\\
$2+$&$1.0$&$0.9$\\
$3+$&$0.9$&$0.7$\\
\end{tabular}
\end{ruledtabular}
\end{table}

\begin{figure}
\includegraphics{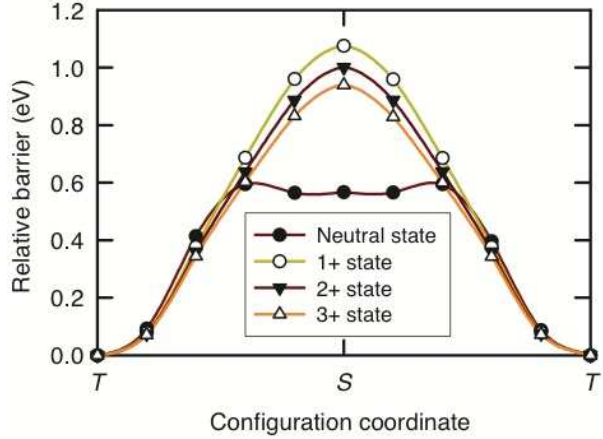}%
\caption{\label{figure:gg2s110s2gg_barrier}
(Color online)
The energy barriers computed with the nudged
elastic band method are shown
for the migration of a gallium tetrahedral interstitial between
gallium atoms ($T$) to the next-nearest neighboring tetrahedral interstitial position,
also between gallium atoms, via the gallium-gallium $\langle 110 \rangle$ split-interstitial
position ($S$), in the neutral, +1, +2, and +3 charge states.   The calculated energy at 
each point along the path is plotted vs. the distance along the path.  The energies
presented are all energies relative to the formation energy calculated for the tetrahedral interstitial 
between gallium atoms in the given charge state.  The calculations
were performed at the points indicated and curves are added to guide the eye.}
\end{figure}

The energy as a
function of configuration coordinate for each
charge state is shown in Fig.\ \ref{figure:gg2s110s2gg_barrier} 
for the complete migration path between two tetrahedral 
interstitial sites surrounded by gallium atoms 
passing through the gallium-gallium $\langle 110 \rangle$ split 
interstitial position.
The intermediate point of the fully converged split-interstitial pathway remained at the
gallium-gallium $\langle 110 \rangle$ split-interstitial configuration, 
which can be seen to be a shallow 
metastable configuration for the neutral charge state.
The metastability is visible as a very shallow dip in the curve at the point corresponding to the
neutral $\langle 110 \rangle$ split-interstitial configuration (see 
Fig.\ \ref{figure:gg2s110s2gg_barrier}).

\begin{figure}
\includegraphics{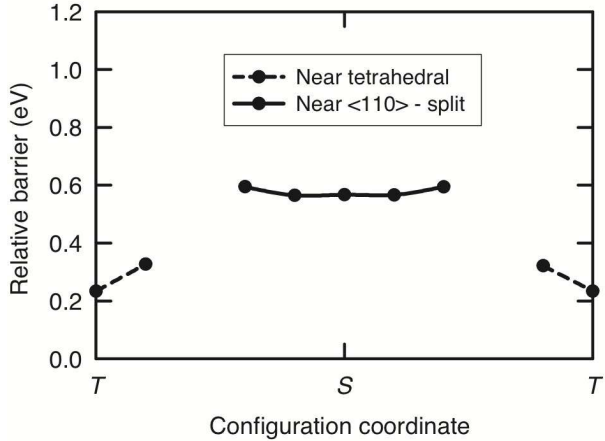}%
\caption{\label{figure:neutralbarrier}
Corrected energy barriers at 300K are shown
for the migration of a neutral gallium tetrahedral interstitial between
gallium atoms ($T$) to the next-nearest neighboring tetrahedral interstitial position,
also between gallium atoms, via the gallium-gallium $\langle 110 \rangle$ split-interstitial
position ($S$).   The energy at 
each point along the path is plotted vs. the distance along the path.  The energies
presented are derived from the nudged 
elastic band results for the same migration presented
in Fig.\ \ref{figure:gg2s110s2gg_barrier}, with a shift of +0.236~eV applied to the
neutral tetrahedral part of the calculation, as discussed in the text.}
\end{figure}

We find that
the energy barrier to leave the neutral gallium-gallium $\langle 110 \rangle$ split 
configuration and continue along the diffusion path is only 0.02 eV,  
making this defect only barely metastable.  This result is in agreement with 
Schultz \emph{et al.},\cite{Schultz2009a} who find that there is (almost) no barrier for the 
neutral gallium-gallium $\langle 110 \rangle$ split interstitial to convert to 
a neutral tetrahedral interstitial surrounded by gallium atoms, and it is  
in qualitative
agreement with Malouin \emph{et al.},\cite{Malouin2007a} who do not find 
the neutral gallium-gallium $\langle 110 \rangle$ split
interstitial to be metastable at all.  

For the +1, +2, and +3 charge states, we find that the gallium-gallium $\langle 110 \rangle$ split 
interstitial configuration is a saddle point at the midpoint of the diffusion path, 
as can be seen in Fig.\ \ref{figure:gg2s110s2gg_barrier}, and 
is therefore unstable. 
In a slightly different result, Schultz 
\emph{et al.}\cite{Schultz2009a}
found enough of a region of metastability to calculate formation energies 
for the gallium-gallium $\langle 110 \rangle$ split 
interstitial configuration in these charge states without doing nudged elastic band calculations 
to investigate diffusion paths that pass through or close to this configuration. 
 
Our results for the energy barriers along this diffusion path for the +1, +2, and +3 charge 
states, which are just 
the differences in formation energy between the gallium-gallium $\langle 110 \rangle$ split 
interstitial and the tetrahedral 
interstitial surrounded by gallium atoms, are consistent with  
the results of Schultz \emph{et al}.  
Schultz \emph{et al}.\ report the formation energy of the 
gallium-gallium $\langle 110 \rangle$ split 
interstitial in the +1 charge state to be 1.12(1.07) eV higher than the formation 
energy of the tetrahedral interstitial surrounded by gallium atoms, based on their 
LDA(PBE) calculations.\cite{Schultz2009a}  This is in good agreement with our 
calculated energy barrier of 1.1 eV 
for gallium interstitial diffusion in the +1 charge state 
between two tetrahedral interstitials surrounded by gallium atoms 
via the path through the 
gallium-gallium $\langle 110 \rangle$ split 
interstitial configuration.
Schultz \emph{et al.}\ report the formation energy 
of the gallium-gallium $\langle 110 \rangle$ split 
interstitial in the +2 and +3 charge states to be 0.9-1.0 eV higher than the formation 
energy of the tetrahedral interstitial surrounded by arsenic atoms.\cite{Schultz2009a}  Although 
Schultz \emph{et al}.\ do not report their results for the +2 and +3 charge states 
in as much detail or with as much precision 
as they do for the +1 charge state, we may still make a comparison 
to our results for the +2 and +3 charge states, by using our results 
that for these charge states, the formation energies of the two tetrahedral interstitials 
are very similar, as shown in Fig.\ \ref{figure:enthalpy} and discussed in 
Section \ref{subsection:energyandstructure}.  We conclude that the formation energy 
differences reported by 
Schultz \emph{et al.}\ are in reasonable agreement with our calculated energy barriers of 
1.0 eV and 0.9 eV for diffusion along this path in the +2 and +3 charge states, respectively.  

Since the energy has a simpler dependence on the configuration coordinate for the 
migration path through the hexagonal 
interstitial configuration, a figure equivalent to 
Fig.\ \ref{figure:gg2s110s2gg_barrier} is not shown for this path.  
The intermediate point of the fully converged hexagonal pathway between 
a tetrahedral interstitial site surrounded by gallium atoms and 
a neighboring tetrahedral interstitial site surrounded by arsenic atoms remained at the
hexagonal interstitial configuration.
This diffusion path has a single peak in energy at the hexagonal configuration, which is a 
saddle point for all charge states.  The final energy is slightly higher 
than the initial energy for this migration path, since the gallium begins in the slightly more 
energetically favorable of the two tetrahedral interstitial configurations and ends in the 
slightly higher energy configuration.  The second step in the migration of a gallium 
atom via the hexagonal path to a tetrahedral site equivalent 
to the starting site would be for it to reverse this process, moving from the slightly higher energy 
tetrahedral interstitial surrounded by arsenic atoms to a 
neighboring tetrahedral interstitial surrounded by 
gallium atoms.

As we have discussed in Section \ref{subsection:energyandstructure}, the 
calculated formation energy for any neutral defect with its last electron occupying a shallow 
hydrogenic bound state composed of conduction band states should be shifted up 
by a scissors-type correction\cite{Baraff1984,Johnson1998a} in order to 
correctly represent a transition level just below the conduction band edge for the 
transition from singly positive to neutral.  Therefore, as indicated in 
Section \ref{subsection:energyandstructure}, 
in our calculations of 
activation energies for diffusion, we shift the formation energy 
for both neutral tetrahedral interstitials and for the neutral hexagonal interstitial 
up by the same scissors correction 
as is needed to shift the effective conduction band edge in our 
defect calculations up to coincide with the 
experimental conduction band edge.  This correction is 0.236~eV at 300K.

Both the initial and final points as well as the saddle point configuration 
for the hexagonal path are all shifted up in energy by the same amount for diffusion 
in the neutral state.  Therefore the calculated barrier shown 
in Table \ref{table:migrationenergies} 
for diffusion of a neutral gallium interstitial via the hexagonal path is not affected by 
the scissors correction.  However, under conditions 
where interstitial concentrations are determined by thermal equilibrium, 
the activation energy for diffusion 
consists of the formation energy plus the 
migration barrier for the interstitial.  Since the formation energy of the initial neutral 
tetrahedral interstitial 
is shifted up, under conditions 
where interstitial concentrations are determined by thermal equilibrium, 
the activation energy for neutral gallium interstitial diffusion 
via the hexagonal 
path will be increased by the scissors correction.

In the case of neutral interstitial diffusion via the gallium-gallium 
$\langle 110 \rangle$ split-interstitial 
path, the scissors correction affects both the formation energy 
of the initial tetrahedral interstitial and the diffusion barrier.  Since 
the last electron bound to the 
neutral gallium-gallium $\langle 110 \rangle$ split interstitial occupies 
a true localized defect state 
rather than a shallow hydrogenic bound state, the formation energy of this neutral 
defect configuration should not be shifted up by the scissors correction.  
Therefore the formation energy of the neutral tetrahedral interstitial configurations which 
form the initial and final points on this path will be shifted up, 
while the formation energy of the neutral $\langle 110 
\rangle$ split interstitial configuration which 
forms the midpoint of this path will not.  This results in an upward shift of the 
energy at the initial and final points of this path, relative to the midpoint, 
by 0.236 eV at 300K.
  
Fig.\ \ref{figure:neutralbarrier} shows the corrected energy as a
function of configuration coordinate for gallium diffusion via 
the gallium-gallium $\langle 110 \rangle$ split 
interstitial pathway in the neutral charge state.  The energies are obtained from the nudged 
elastic band results for the same migration presented
in Fig.\ \ref{figure:gg2s110s2gg_barrier}, with a shift of +0.236~eV applied to the
neutral tetrahedral interstitial part of the calculation.   We have only included points on this
graph near the two endpoints and the midpoint --- however, there is no evidence to suggest 
that any more significant additional barrier will develop around the $\langle 110 \rangle$ split 
interstitial as a result of including corrections to the formation energy of the 
shallow hydrogenic bound state for the initial and final tetrahedral interstitial configurations.  
We conclude that the corrected barrier, defined as the difference in energy between the initial 
configuration and the highest energy configuration along the path, is about 0.35 eV.  

For neutral gallium diffusion via the in-network path through the 
gallium-gallium $\langle 110 \rangle$ split 
interstitial configuration, the activation energy for diffusion 
under conditions of thermal equilibrium 
consists of the formation energy of the initial interstitial configuration 
(which is shifted up due to the scissors correction) plus the 
migration barrier (which is shifted down by the same amount).  Therefore the activation energy 
under conditions of thermal 
equilibrium  for gallium diffusion 
in the neutral state via the gallium-gallium $\langle 110 \rangle$ split 
interstitial 
pathway will not be affected by the scissors correction.  
In fact, the activation energy under conditions of thermal 
equilibrium  for interstitial diffusion via any path is merely the formation 
energy of the interstitial configuration at the highest energy point 
in the path --- shifting the 
formation energy of the initial configuration up or down relative to the 
highest energy configuration 
does not affect this activation energy.

When we compare the energy barriers for gallium interstitial diffusion 
through the hexagonal configuration and for 
gallium interstitial diffusion 
through the gallium-gallium $\langle 110 \rangle$ split 
interstitial configuration, we find that the charge state 
of the diffusing interstitial determines which pathway has the lowest barrier. 
Calculated diffusion barriers are lower for diffusion of the excess gallium atom 
through the hexagonal configuration for the $q = +2$ and $+3$ charge states, and   
diffusion barriers are lower for the in-network path involving gallium interstitial diffusion 
through the gallium-gallium $\langle 110 \rangle$ split 
interstitial configuration for the neutral and +1 charge states.  
This suggests that the Fermi level is likely to play a role in 
determining the relative importance of 
these two diffusion pathways.  
 
Gallium interstitial diffusion may also proceed through other in-network paths.   
However, we expect that the dominant contributions to diffusion will be due to the 
in-network path through the gallium-gallium $\langle 110 \rangle$ split 
interstitial configuration and the non-network path through the hexagonal site.  
For the +2 and +3 charge states, 
the formation energies for 
all the other metastable split interstitials which have been 
studied\cite{Malouin2007a,Schultz2009a} are at least 0.9 eV higher that 
the formation energy of the tetrahedral 
interstitial surrounded by gallium atoms.  Any additional energy needed 
to move into and out of these configurations would further raise the energy barrier 
for diffusion through these configurations.  Since the entire energy barrier for diffusion 
through the hexagonal site is 0.9 eV for the +2 charge state and 0.7 eV for the +3 charge state, 
according to our calculations, we expect this to be the lowest energy pathway 
(or at least one of the lowest energy pathways) for diffusion in the +2 and +3 charge states.
For the 
neutral interstitial, the lowest energy non-tetrahedral configuration which has been 
studied is the gallium-gallium $\langle 110 \rangle$ split 
interstitial, and there is a negligible additional barrier 
for moving into and out of this configuration, as discussed above and in
Section \ref{subsection:energyandstructure}.  We conclude that the lowest energy 
pathway for diffusion in the neutral state will be the pathway through the 
gallium-gallium $\langle 110 \rangle$ split 
interstitial.  For the +1 charge state, diffusion may proceed through the 
gallium-arsenic 
$\langle 111 \rangle$ split interstitial configuration, with a barrier of 1.4 eV, as calculated by 
Levasseur-Smith \emph{et al.}\cite{Levasseur-Smith2008a} and discussed above, as well as through 
the gallium-gallium $\langle 110 \rangle$ split 
interstitial configuration, with a barrier of 1.1 eV, as we have calculated above.  
We conclude that although both pathways may contribute to diffusion in the +1 charge state, 
the pathway through 
the gallium-gallium $\langle 110 \rangle$ split 
interstitial configuration appears to be the lower energy pathway.

\subsection{\label{subsection:dependence}Diffusion dependence on stoichiometry and doping}

In this section, we combine the calculated defect formation energies and 
migration barriers in order to get an overall picture of the contribution 
of gallium interstitials to
gallium diffusion in gallium arsenide under different experimental conditions.  
Because of the dependence of the dominant charge states 
and the formation energies of defects on the Fermi energy
and chemical potential, we examine the contributions 
due to different defect pathways and different charge states to
gallium diffusion as a function of these experimental conditions.

Changes in the chemical potential raise or lower the formation energies of 
all gallium interstitial configurations by the same amount, since they all involve 
one excess gallium atom.  
We have presented the lower and upper limits for the formation energies of gallium 
interstitials, corresponding to the 
gallium-rich and arsenic-rich limits, respectively, as a function of the Fermi level in
Figs.\ \ref{figure:enthalpy} and \ref{figure:asrichenthalpy}.  Under conditions 
where interstitial concentrations are determined by thermal equilibrium, changes in 
the chemical potential will raise or lower the activation 
energies by the same amount for gallium interstitial diffusion in all charge 
states and by all pathways.  However, changes in the chemical potential do 
not change the relative importance of different diffusion pathways for the interstitial 
or diffusion in diffent charge states.  In the following discussion of the 
effect of the Fermi level on gallium interstitial diffusion, 
we will focus on diffusion for a chemical potential in the gallium-rich limit. 

The experimentally measured Fermi level is 
determined by many factors, including concentrations of native defects 
produced by thermal equilibrium or 
by processes such as irradiation or ion implantation, as well as 
intentional or unintentional doping.  
For example, as shown in Section \ref{subsection:equilibriumdefectconcentrations}, 
changes in the chemical potential affect the relative concentrations 
in thermal equilibrium of native defects containing different 
numbers of excess gallium or arsenic atoms, many of which are charged.  Therefore changes 
in the chemical potential may 
contribute to changes in the Fermi level.
 
\begin{figure}
\includegraphics{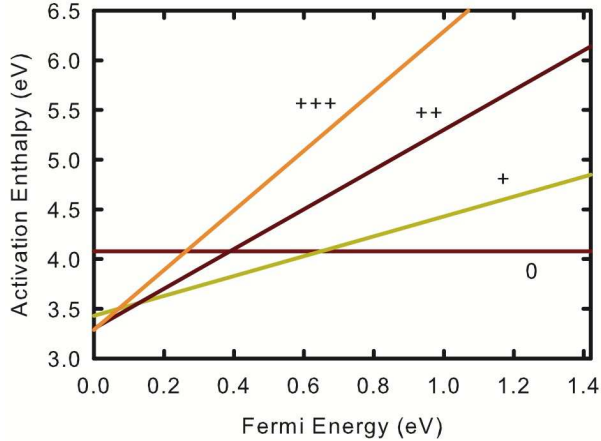}%
\caption{\label{figure:tet2split110createandmigrate}
(Color online) The activation enthalpy for diffusion through the 
$\langle 110 \rangle$ gallium-gallium split interstitial configuration 
for charge states from neutral to +3, as a function of Fermi energy $\epsilon_F$, 
in the gallium-rich limit, under conditions 
where interstitial concentrations are determined by thermal equilibrium.}
\end{figure}

The dependence of gallium interstitial diffusion on Fermi level is more complicated 
than its dependence on the chemical potential.
In Fig.\ \ref{figure:tet2split110createandmigrate} we display the
activation enthalpy for
gallium interstitial diffusion in the gallium-rich limit as a function of
the Fermi energy for diffusion through 
the gallium-gallium $\langle 110 \rangle$ split 
interstitial configuration, under conditions 
where interstitial concentrations are determined by thermal equilibrium.  
Activation enthalpies are shown for diffusion via this path 
for interstitials
in each of the four charge states from neutral to triply positive.  In this figure 
as well as all the other figures in this section, 
results for 300\,K are shown: the  
Fermi energy is shown varying across the 300\,K experimental band gap
of 1.42~eV.  

The activation enthalpy for gallium interstitial diffusion 
shown in Fig.\ \ref{figure:tet2split110createandmigrate}  
is the energy required for 
a gallium tetrahedral interstitial surrounded by gallium atoms to be created
and then to migrate through the $\langle 110 \rangle$ gallium-gallium split
interstitial position to the next equivalent tetrahedral 
interstitial position.  Fig.\ \ref{figure:tet2split110createandmigrate} shows 
the activation enthalpy obtained from our \textsc{vasp} calculations, 
which is the sum of the initial interstitial formation energy and the migration barrier.
As discussed in Section \ref{subsection:pathways for diffusion}, this activation 
enthalpy is equal to the 
enthalpy of formation of the defect at the highest energy point 
along the path --- in this case, 
the formation energy of the $\langle 110 \rangle$ gallium-gallium split
interstitial.  Therefore any uncertainty in the formation energy 
of the tetrahedral interstitial has no effect on this graph.  

Since the activation enthalpy for diffusion via the 
$\langle 110 \rangle$ gallium-gallium split
interstitial path is equal to the 
formation energy of the $\langle 110 \rangle$ gallium-gallium split
interstitial, we may compare our results for this formation energy as obtained 
in our \textsc{vasp} and our norm-conserving pseudopotential calculations, to see whether 
these calculations lead to similar results.  Fig.\ \ref{figure:enthalpy} 
in Section \ref{subsection:energyandstructure} shows the results of our
norm-conserving pseudopotential calculations for the 
formation energy of the $\langle 110 \rangle$ gallium-gallium split
interstitial in its most energetically favorable charge state in the 
gallium-rich limit, as a 
function of Fermi energy across the calculated band gap.  Although the scissors 
correction has not been applied to raise the band gap in this figure 
to its experimental value at 300K, the results of our norm-conserving pseudopotential 
calculations shown in Fig.\ \ref{figure:enthalpy} may still be compared to the 
results of our \textsc{vasp} calculations over the range of Fermi levels corresponding 
to the calculated band gap.  
The results of our norm-conserving pseudopotential 
calculations for the 
formation energy of the $\langle 110 \rangle$ gallium-gallium split
interstitial in its most energetically favorable charge state, 
shown in Fig.\ \ref{figure:enthalpy}, are essentially the same as the results 
of our \textsc{vasp} calculations for the activation enthalpy in the most energetically 
favorable charge state, shown in Fig.\ \ref{figure:tet2split110createandmigrate}.

As we can see in 
Fig.\ \ref{figure:tet2split110createandmigrate}, interstitial diffusion through the 
$\langle 110 \rangle$ gallium-gallium split
interstitial pathway is largely dominated by diffusion in the neutral and +1 states. The
the activation enthalpy for diffusion via this pathway is lowest for 
diffusion in the neutral and +1 charge states 
across much of the available range of Fermi energies.  Only in highly $p$-doped material, 
with a Fermi level approaching the valence band edge, are the activation 
enthalpies for diffusion in 
the +2 and +3 charge
states as low or lower than the activation enthalpy for diffusion in the +1 state.

\begin{figure}
\includegraphics{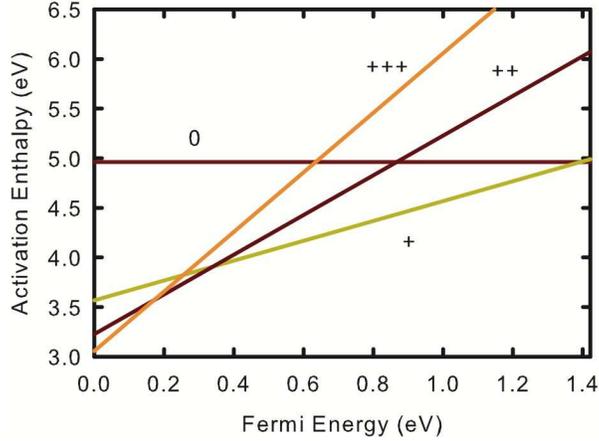}%
\caption{\label{figure:createandmigrate}
(Color online) The activation enthalpy for diffusion through the 
hexagonal interstitial position 
for charge states from neutral to +3, as a function of Fermi energy $\epsilon_F$, 
in the gallium-rich limit, under conditions 
where interstitial concentrations are determined by thermal equilibrium.}
\end{figure}

In Fig.\ \ref{figure:createandmigrate} we display a similar graph for the case in which
the tetrahedral gallium interstitial migrates from its location between gallium atoms to the
nearest tetrahedral location between arsenic atoms via the hexagonal opening
in the lattice between the two tetrahedral locations.  Again, we note that 
the activation enthalpy presented in this graph is just the 
formation energy of the defect at the highest energy point 
along the path.  The highest energy point along the path shown in 
Fig.\ \ref{figure:createandmigrate} corresponds to the hexagonal interstitial; 
therefore changes in the formation energy of the initial and final tetrahedral 
interstitials do not affect this graph.  

Since the activation enthalpy for diffusion via the 
hexagonal interstitial path is equal to the 
formation energy of the hexagonal 
interstitial, we may compare the results of our norm-conserving pseudopotential 
calculations for the formation energy of the 
hexagonal interstitial to the results of our \textsc{vasp} calculations for the activation enthalpy for 
gallium interstitial diffusion through 
the hexagonal position.  Fig.\ \ref{figure:enthalpy} 
in Section \ref{subsection:energyandstructure}
shows the results of our norm-conserving pseudopotential 
calculations for the 
formation energy of the hexagonal
interstitial in its most energetically favorable charge state in the 
gallium-rich limit, as a 
function of Fermi energy across the calculated band gap.  
These results are essentially the same as the results 
of our \textsc{vasp} calculations for the activation enthalpy in the most energetically 
favorable charge state, shown in Fig.\ \ref{figure:createandmigrate}.

Comparison of Figs.\ \ref{figure:tet2split110createandmigrate} 
and \ref{figure:createandmigrate} suggests that diffusion can be modeled 
more simply for high and mid-gap Fermi energies than it can be for 
Fermi energies near the valence band edge.  These figures show that 
the neutral and +1 charge states are the 
most energetically favorable charge states for 
diffusion via the 
split interstitial 
and hexagonal pathways over
most of the range of possible Fermi energies, except for 
Fermi energies near the valence band edge.  We also observe that activation energies 
are lower for diffusion via the split interstitial pathway than for 
diffusion via the hexagonal pathway for both the neutral and the +1 charge states.  
We conclude that diffusion proceeds primarily through the in-network path through the 
$\langle 110 \rangle$ gallium-gallium split interstitial configuration in 
the neutral and +1 charge states over
most of the range of possible Fermi energies, except for 
Fermi energies near the valence band edge.  Diffusion is 
most energetically favorable in the neutral charge state for Fermi 
energies around midgap and above.
Diffusion is 
most energetically favorable in the +1 charge state for Fermi energies between 
mid-gap and the region near the valence band edge.  For Fermi energies 
where the +1 charge state dominates diffusion, overall diffusion 
may be enhanced by small additional contributions 
to diffusion in this charge state by the hexagonal pathway and the pathway through the 
$\langle 111 \rangle$ gallium-arsenic split interstitial configuration, which was explored by 
Levasseur-Smith \emph{et al.}\cite{Levasseur-Smith2008a} and discussed above in 
Section \ref{subsection:pathways for diffusion}.  

We find that gallium interstitial diffusion should be the 
most complicated to model and also the most rapid for Fermi energies near 
the valence band edge, 
when concentrations of interstitials are determined by thermal equilibrium. 
Figs.\ \ref{figure:tet2split110createandmigrate} 
and \ref{figure:createandmigrate} show that  
the activation
enthalpies in the most energetically favorable charge state 
for both the $\langle 110 \rangle$ gallium-gallium split
interstitial path and the hexagonal path are lowest 
when the Fermi energy is low.  This should result in more rapid diffusion 
via these paths under conditions of $p$-type doping.  
The most energetically favorable diffusion for Fermi energies below about 0.2 eV is 
diffusion in the +2 or +3 charge state via the hexagonal pathway.  
However, no single charge state can be said to dominate diffusion 
for a Fermi energy with 0.2~eV of the valence band edge, as 
activation energies for diffusion in the +1 and +2 charge states are very similar 
at the upper end of this range, and activation energies for diffusion in the 
+2 and +3 charge states are very similar as the Fermi energy approaches the 
valence band edge.  This can lead to a complicated 
diffusion profile, and difficulties extending models for diffusion which have 
worked well for higher Fermi energies into this region.

\begin{figure}
\includegraphics{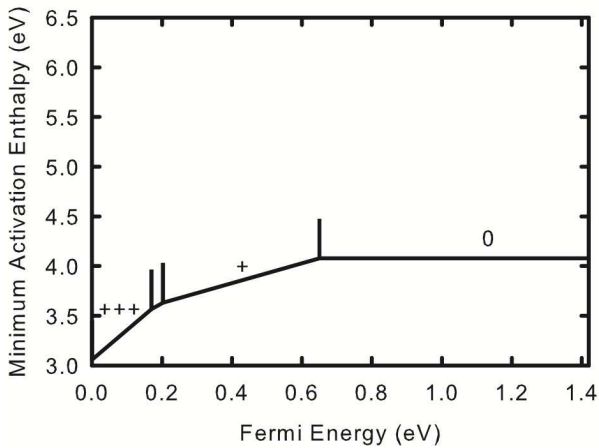}%
\caption{\label{figure:minimumactivation}
The minimum activation enthalpy for gallium interstitial diffusion via the pathways
calculated in this paper as a function of Fermi energy $\epsilon_F$, 
in the gallium-rich limit, under conditions 
where interstitial concentrations are determined by thermal equilibrium.
In the +2 and +3 charge states, migration through
the hexagonal configuration possesses the lowest activation enthalpy.
In the neutral and +1 charge states, migration through
the $\langle 110 \rangle$
split interstitial configuration possesses the lowest activation enthalpy.}
\end{figure}

Fig.\ \ref{figure:minimumactivation} shows the lowest
activation enthalpies for diffusion via either the 
hexagonal path or the $\langle 110 \rangle$ gallium-gallium split
interstitial path in the gallium-rich limit, as a function of the 
Fermi energy across the experimental 300\,K band gap, for diffusion 
when 
interstitial concentrations are determined by thermal equilibrium.  
In the +2 and 
+3 charge states, the lowest activation enthalpy involves diffusion via the hexagonal
defect pathway.  In the +1 and neutral 
charge states, the lowest activation enthalpy involves diffusion via the split
interstitial pathway.  As shown in Fig.\ \ref{figure:minimumactivation}, the 
lowest activation enthalpy for diffusion when 
interstitial concentrations are determined by thermal equilibrium is  
3.1 eV in the gallium-rich limit, corresponding to 
diffusion via the hexagonal 
path in the +3 charge state.  Under arsenic-rich conditions, due to the larger 
energy cost for forming gallium interstitials, the activation enthalpies at 
all Fermi energies are increased by 0.5 eV.     

\begin{figure}
\includegraphics{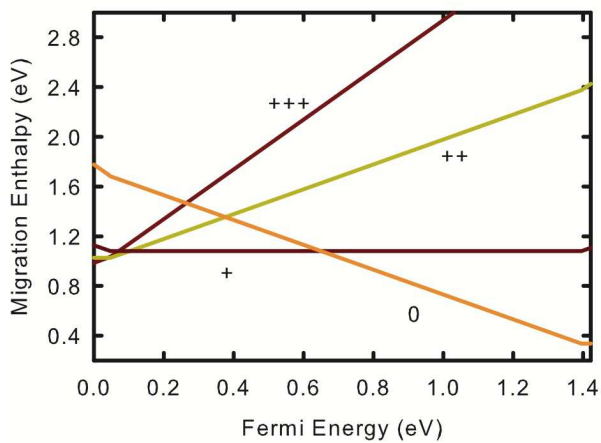}%
\caption{\label{figure:tet2split110changeandmigrate}
(Color online) The 
energy needed for a pre-existing
gallium tetrahedral interstitial between Ga atoms to change from the 
minimum energy charge state to the charge state
specified and then migrate to another tetrahedral site, passing through
the $\langle 110 \rangle$ gallium-gallium split interstitial configuration,
is plotted as a function of $\epsilon _F$.}
\end{figure}

\begin{figure}
\includegraphics{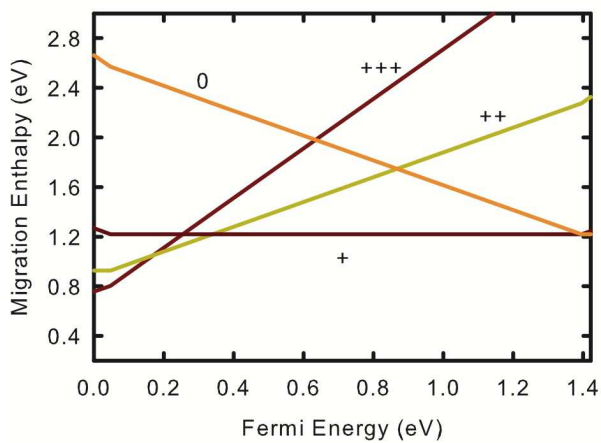}%
\caption{\label{figure:changeandmigrate}
(Color online) The 
energy needed for a pre-existing
gallium tetrahedral interstitial between Ga atoms to change from the 
minimum energy charge state to the charge state
specified and then migrate to another tetrahedral site, passing through
the hexagonal gallium interstitial configuration,
is plotted as a function of $\epsilon _F$.}
\end{figure}

If gallium interstitials have been formed through damage associated with implantation,
radiation, or processing, then we may consider migration starting
from an abundance of non-equilibrium interstitials in the material.
These pre-existing interstitials are mostly present in their lowest energy charge state, which
depends on the Fermi energy, but they may diffuse primarily in a different charge state which
has a lower migration barrier.
In this case, the activation energy for diffusion to occur in a given charge state is
the energy it takes to promote the pre-existing defect to this charge state
from the lowest energy charge state, plus the migration barrier for the given charge state.  
These activation energies 
are plotted as a function of
Fermi energy in
Fig.\ \ref{figure:tet2split110changeandmigrate} and
Fig.\ \ref{figure:changeandmigrate} for the $\langle 110 \rangle$ 
gallium-gallium split interstitial pathway and
and the hexagonal pathway, respectively.  

As can be seen by comparing Figs.\ \ref{figure:tet2split110changeandmigrate} and
\ref{figure:changeandmigrate} to the corresponding figures for diffusion when 
interstitial concentrations are 
determined by thermal equilibrium, Figs.\ \ref{figure:tet2split110createandmigrate} 
and \ref{figure:createandmigrate}, the presence of non-equilibrium concentrations of 
interstitials does not affect which pathway and charge state has the lowest 
activation energy for diffusion for a given Fermi energy.  Regardless of 
whether we consider the existing interstitial concentrations to be in thermal equilibrium or out
of equilibrium, 
we conclude that diffusion proceeds primarily via the in-network path through the 
$\langle 110 \rangle$ gallium-gallium split interstitial configuration in 
the neutral charge state for Fermi 
energies around midgap and above, and via the same path in the +1 charge 
state for Fermi energies between 
mid-gap and the region near the valence band edge.  
Similarly, we conclude that diffusion via the hexagonal path in the +2 and +3 charge states 
becomes the most energetically favorable for Fermi energies within 0.2 eV 
of the valence band edge.  As in the case when interstitial concentrations are 
determined by thermal equilibrium, 
no single charge state can be said to dominate diffusion 
for a Fermi energy within 0.2~eV of the valence band edge, as 
activation energies for diffusion in the +1 and +2 charge states are very similar 
at the upper end of this range, and activation energies for diffusion in the 
+2 and +3 charge states are very similar as the Fermi energy approaches the 
valence band edge.   

However the activation 
energy for interstitial diffusion is larger 
when non-equilibrium concentrations of pre-existing interstitials are not available.  
The activation energy for diffusion when 
interstitial concentrations are 
determined by thermal equilibrium is equal to the activation energy for diffusion 
when there are 
non-equilibrium concentrations of pre-existing interstitials plus the additional formation 
energy needed to create 
an interstitial in its lowest energy configuration and charge state at the given Fermi energy.

\begin{figure}
\includegraphics{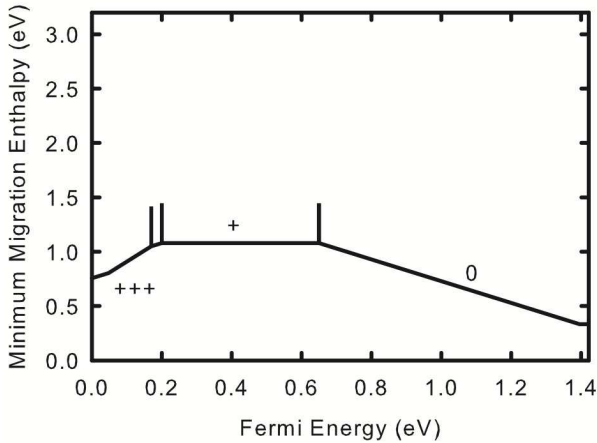}%
\caption{\label{figure:minimumactivationpreexistinginterst}
The minimum activation enthalpy for gallium interstitial diffusion via the pathways
calculated in this paper as a function of Fermi energy $\epsilon_F$, 
in the gallium-rich limit, under conditions 
where there are non-equilibrium concentrations of pre-existing interstitials.  
In the +2 and +3 charge states, migration through
the hexagonal configuration possesses the lowest activation enthalpy.
In the neutral and +1 charge states, migration through
the $\langle 110 \rangle$
split interstitial configuration possesses the lowest activation enthalpy.}
\end{figure}

Fig.\ \ref{figure:minimumactivationpreexistinginterst} shows the lowest
activation enthalpies for diffusion via either the 
hexagonal path or the $\langle 110 \rangle$ gallium-gallium split
interstitial path in the gallium-rich limit, as a function of the 
Fermi energy across the experimental 300\,K band gap, for diffusion when there are 
non-equilibrium concentrations of pre-existing interstitials.  
As for the case when 
interstitial concentrations are determined by thermal equilibrium, in the +2 and 
+3 charge states, the lowest activation enthalpy involves diffusion via the hexagonal
defect pathway.  In the +1 and neutral 
charge states, the lowest activation enthalpy involves diffusion via the split
interstitial pathway.  

We conclude from the results shown in Fig.\ \ref{figure:minimumactivationpreexistinginterst} 
that if there are non-equilibrium concentrations of 
pre-existing interstitials, gallium interstitial diffusion will be somewhat enhanced by 
strong $p$-type doping, and greatly enhanced by strong $n$-type doping.   
As in the case where interstitial concentrations 
are determined by thermal equilibrium, 
we see in Fig.\ \ref{figure:minimumactivationpreexistinginterst} 
that the activation enthalpy for diffusion is reduced as 
the Fermi level approaches the valence band edge.  This is due to the low barrier 
of 0.7 eV for diffusion via the hexagonal path in the +3 charge state, which should result in  
enhanced diffusion when lower Fermi levels make it more likely for any 
existing interstitials to be present in this charge state.  However, in contrast to the 
behavior we see when interstitial concentrations 
are determined by thermal equilibrium, if there are non-equilibrium concentrations of 
pre-existing interstitials, we see that the activation enthalpy for diffusion is reduced 
considerably more as 
the Fermi level approaches the conduction band edge.  This is due to the even lower barrier 
of 0.35 eV for diffusion via the split interstitial path in the neutral charge state, 
which should result in greatly  
enhanced gallium interstitial diffusion when Fermi levels approaching the conduction band edge 
make it more likely for any 
existing interstitials to be present in the neutral charge state.

Comparison to experimental results requires that we take into account the fact 
that both gallium interstitials and gallium vacancies may contribute to 
gallium diffusion in gallium arsenide.  
Modeling of experimental results, as reviewed in,\cite{Tan1991a}
has often been used to help identify the dominant 
contributor to gallium diffusion.  
As a result of this modeling, the gallium vacancy has been identified as the most 
likely agent for gallium diffusion
for intrinsic and $n$-type doping, while  
the gallium interstitial  has been identified as the 
most likely agent for gallium diffusion for heavy $p$-type doping.\cite{Tan1991a}
To determine the circumstances under which interstitials or vacancies dominate
gallium diffusion theoretically, 
we must compare the activation enthalpies  
across the ranges of doping from $p$-type to $n$-type and composition from
gallium rich to arsenic rich.  

For the contributions to gallium diffusion 
due to gallium vacancies, we will focus on diffusion involving 
the simple gallium vacancy.  A more complicated equivalent defect, consisting of 
a complex of an arsenic vacancy and an 
arsenic antisite, has been identified as having a lower formation energy than 
the simple gallium vacancy for Fermi energies less than 0.3\,eV above the
valence band edge.\cite{Bockstedte1997b,Schultz2009a}  However it is difficult 
to envision a low-energy gallium diffusion pathway 
which primarily involves this complex.  A second-neighbor hop of a diffusing gallium 
atom from a gallium lattice site into a gallium vacancy has the net effect of 
moving the gallium vacancy 
in the opposite direction, without creating any further defects.  However, if a 
diffusing gallium atom moves into the arsenic vacancy part 
of the arsenic vacancy-arsenic antisite complex, 
this produces a complex consisting of a pair of antisites.  Further 
movement of the entire Ga$_\mathrm{As}$-As$_\mathrm{Ga}$
complex would not lead to any further net motion of gallium.  Even in strongly
$p$-type gallium arsenide, where the arsenic vacancy-arsenic antisite complex should 
be somewhat more numerous 
than the simple gallium vacancy, the average distance between these complexes will be 
considerably further than a second-neighbor distance.  Movement of a diffusing gallium 
atom between the arsenic vacancies in these complexes would require the diffusing atom to 
come out of the vacancy and diffuse a significant distance as an interstitial before 
finding another arsenic or gallium vacancy.

There is basic agreement on the general picture of gallium vacancies obtained from 
theoretical calculations and experimental results.   Our calculated 
formation energies 
for simple gallium vacancies appear in Figs.\ \ref{figure:enthalpy}
and \ref{figure:asrichenthalpy} for the gallium-rich limit and 
the arsenic-rich limit, respectively.  
We find that the energetically preferred charge 
states of the simple gallium vacancy range from 
neutral to $-3$, as a function of the Fermi level, in general agreement with previous 
calculations.\cite{Baraff1985a,Northrup1993a,Schick2002a,Gorczyca2002a,
Janotti2003a,ElMellouhi2005a,Schultz2009a}  
In agreement with this theoretical picture, charge states of the gallium vacancy 
which have been used  
in the modeling of gallium diffusion 
range from triply 
negative for $n$-type
to singly negative for $p$-type.\cite{Tan1991a}  
A positron annihilation experiment is consistent with the
gallium vacancy existing in the triply negative charge state in $n$-type
material.\cite{Gebauer2001b}

Energy barriers for diffusion of the gallium vacancy have been
calculated by two theoretical groups.  Bockstedte and Scheffler\cite{Bockstedte1997b}
identified the diffusion pathway as a second neighbor hop, with a barrier
of 1.7~eV in the neutral charge state and 1.9~eV in the triply negative charge state.
Subsequently El-Mellouhi and Mousseau\cite{ElMellouhi2006a} identified another second-neighbor
hop (different in the details) to be the lowest-energy pathway 
for gallium vacancy diffusion, with energy
barriers of 1.7, 1.7, 1.84, and 2.0~eV in the 
neutral, singly negative, doubly negative, 
and triply negative charge states, respectively.  As discussed above, these are the 
most energetically favorable 
charge states of the simple gallium vacancy for a Fermi level anywhere in the gap.  
We have used the energy barriers for 
gallium vacancy diffusion calculated by El-Mellouhi and Mousseau\cite{ElMellouhi2006a} 
when calculating activation enthalpies for gallium vacancy diffusion.

All the calculated diffusion barriers for gallium vacancies are larger 
than all the calculated diffusion barriers for gallium 
interstitials, for defects in 
any of the dominant charge 
states across the possible range of 
Fermi energies.  Therefore activation enthalpies for diffusion 
will be larger for pre-existing, non-equilibrium concentrations of vacancies than 
for similar pre-existing concentrations of interstitials. 
Irradiation may knock gallium atoms far enough from 
their lattice sites to leave behind separate vacancy-rich and interstitial-rich regions.  
Due to the lower activation enthalpies for diffusion of pre-existing gallium 
interstitials, we may expect 
more rapid gallium diffusion in the interstitial-rich regions than in 
the vacancy-rich regions in irradiated GaAs.

For the case when concentrations of defects 
are determined by thermal equilibrium, we calculated the activation
enthalpies for gallium vacancy diffusion using our numbers for formation energies
and the barriers calculated by 
El-Mellouhi and Mousseau,\cite{ElMellouhi2006a} listed above.  
We find that the vacancy will be the dominant agent in gallium diffusion
for $n$-type gallium arsenide 
across the entire range of chemical potentials from gallium-rich to arsenic-rich.  
For example, for a Fermi level 0.8~eV above the valence band edge, 
the activation enthalpy for vacancy diffusion in the most favorable charge state 
is 0.9~eV (1.9~eV) lower than 
the activation enthalpy for interstitial diffusion in the most favorable charge state 
in the gallium-rich (arsenic-rich) limit.  The charge state
of the gallium vacancy which dominates diffusion under these conditions is triply negative.

For $p$-type GaAs with defect concentrations determined by thermal equilibrium, we find 
that the relative contributions of vacancies and interstitials to gallium diffusion 
are more dependent on the stoichiometry.  For example,  
in the gallium-rich limit, for a Fermi level 0.2~eV 
above the valence band edge, the activation enthalpy for interstitial diffusion 
in the most favorable charge state 
is 1.2~eV lower than 
the activation enthalpy for vacancy diffusion in the most favorable charge state.  
In the arsenic-rich limit, for the same Fermi level,  
the activation enthalpies for interstitial and vacancy diffusion are essentially the same 
(the activation enthalpy is 0.1~eV lower for interstitial diffusion than for vacancy diffusion).  
We conclude that for a Fermi level 0.2~eV 
above the valence band edge, interstitials dominate diffusion 
in gallium-rich gallium arsenide, while 
both interstitials and vacancies make significant contributions to diffusion in 
arsenic-rich gallium arsenide.
  
The complete picture we have developed for gallium diffusion
in gallium arsenide across the full ranges of Fermi level and stoichiometry 
can allow us to resolve an apparent disagreement between the 
models obtained by fitting some of the experiments described in Section
\ref{section:introduction}.   
First, we note that gallium diffusion
in gallium-rich material will involve gallium interstitials in the
the +1 and neutral charge states when near intrinsic doping
conditions and in the +3 and +2 charge states when doping is
heavily $p$-type, according to our results as shown in 
Fig.~\ref{figure:minimumactivation}.    
For example, we may consider an experiment where diffusion occurs 
at an annealing temperature of 1100~K. 
We self-consistently evaluated the defect concentrations as discussed
in Section \ref{subsection:equilibriumdefectconcentrations}, and found that 
the Fermi energy is about 0.6~eV under intrinsic conditions at 1100~K.  
At this Fermi level, we find that 
the +1 charge state of the interstitial will have the lowest activation enthalpy for diffusion, 
with the neutral state being energetically competitive.  

In recent experiments on zinc-gallium co-diffusion into gallium arsenide resulting in low 
concentrations of zinc, the authors conclude from continuum model fits
that the neutral and +1 charge states of gallium interstitials will be the
most significant interstitials involved in gallium diffusion.\cite{Bracht2005a,BrachtPrivate}
Our calculations are consistent
with this conclusion for low 
concentrations of zinc, since 
this material is not 
heavily $p$-doped.  The authors also note that their model was unable to properly fit 
the diffusion profiles 
for gallium-rich gallium arsenide in regions with higher concentrations of 
zinc.\cite{Bracht2005a,BrachtPrivate}  
This is not surprising, since we would expect more highly charged interstitials 
to contribute more significantly than the neutral and +1 charged interstitials and the gallium 
vacancies which were included in the model in gallium-rich, highly $p$-doped material.

Earlier experiments that indicated that the more highly positively charged states (+2 and +3)
of gallium interstitials are the most important charge states 
for diffusion, on the
other hand, were prepared with more significant concentrations of $p$-type
dopants.\cite{Tan1991a,Yu1991b,Bosker1995a,Bosker1999a}
Our calculations support this conclusion as well, because
this material was prepared under the more heavily $p$-doped conditions 
which favor gallium interstitial
diffusion in the +2 or +3 charge states.  The authors of one of these papers 
noted that there were difficulties 
involved in using +2 charged gallium interstitials to properly fit the diffusion profiles 
in the regions with 
low concentrations of dopants, and that the solution to these difficulties 
would probably involve the use of +1 charged interstitials in the model, in 
addition to +2 and +3 charged interstitials.\cite{Bosker1995a}

Taking into account all these experimental results and the results of our calculations,
we see that many details of the preparation of the material must be known in order
to correctly predict which defects and charge states should be included 
when modeling the material.  Local doping, stoichiometry, and any
non-equilibrium defect concentrations all can play a role in determining what 
should go into the model.

\section{\label{section:conclusion}Conclusion}

In this investigation, the formation enthalpies, structural properties,
and activation enthalpies for diffusion of gallium interstitials in gallium arsenide
across the entire range of chemical potentials from the arsenic-rich limit to the
gallium-rich limit and across the range of doping level from $p$-type to $n$-type
were examined.  Activation enthalpies for diffusion were calculated both for conditions where 
interstitial concentrations are determined by thermal equilibrium, and for material 
containing pre-existing, non-equilibrium concentrations of interstitials.  Regions with 
substantial non-equilibrium concentrations of interstitials may be present in material 
which has been irradiated.

From comparisons to published results for energy barriers for diffusion of
gallium vacancies, it was shown that gallium vacancies
in the triply negative charge state will dominate diffusion in $n$-type material,
that both gallium interstitials and gallium vacancies can be
important agents for gallium diffusion in material that is $p$-type and arsenic
rich, and that gallium interstitials in singly, doubly, and triply positive
charge states will dominate diffusion in gallium-rich gallium arsenide with 
various levels of $p$-type doping.

The results presented in this paper demonstrate that the complete
picture for diffusion of gallium in gallium arsenide is sensitive to the
both the stoichiometry and doping of the material and involves a
variety of defects, migration pathways, and charge states across the
accessible ranges of doping and stoichiometry.  The general picure we have obtained can 
offer guidance on which 
defects and charge states should be considered for successful modeling of gallium diffusion, 
and which can be ignored, 
depending on the details of the experiment.  
For example, these results allow confirmation with microscopically based
theory that in bulk modeling of gallium diffusion it is essential to include 
doubly and triply positive gallium interstitials for strongly 
$p$-doped, gallium-rich gallium arsenide; however it may be acceptable only
to consider the singly positive charge state of the gallium interstitial
when the doping level is closer to intrinsic than strongly $p$-doped, 
especially if the material is not gallium-rich.

\begin{acknowledgments}

The authors are grateful for support from the Air Force Office of Scientific Research for grants
of computer time at computing centers located at AFRL, ARL, NAVO, and ERDC under
DOD HPCMO Project No.\ AFOSR11693MO1.  Some of the preliminary work was done
with support from the Air Force Office of Scientific Research under Grants
No.\ F49620-96-1-0167 and F49620-97-1-0479, and additional grants for computer
time at the NSF/NPACI Supercomputing Centers at SDSC and TACC and  on the Wayne
State University Grid.  And one author (JTS) gratefully acknowledges enlightening
discussion with H. Bracht.

\end{acknowledgments}

%


\begin{thebibliography}{10}%
\makeatletter
\providecommand \@ifxundefined [1]{%
 \ifx #1\undefined \expandafter \@firstoftwo
 \else \expandafter \@secondoftwo
\fi
}%
\providecommand \@ifnum [1]{%
 \ifnum #1\expandafter \@firstoftwo
 \else \expandafter \@secondoftwo
\fi
}%
\providecommand \enquote [1]{``#1''}%
\providecommand \bibnamefont  [1]{#1}%
\providecommand \bibfnamefont [1]{#1}%
\providecommand \citenamefont [1]{#1}%
\providecommand\href[0]{\@sanitize\@href}%
\providecommand\@href[1]{\endgroup\@@startlink{#1}\endgroup\@@href}%
\providecommand\@@href[1]{#1\@@endlink}%
\providecommand \@sanitize [0]{\begingroup\catcode`\&12\catcode`\#12\relax}%
\@ifxundefined \pdfoutput {\@firstoftwo}{%
 \@ifnum{\z@=\pdfoutput}{\@firstoftwo}{\@secondoftwo}%
}{%
 \providecommand\@@startlink[1]{\leavevmode}%
 \providecommand\@@endlink[0]{}%
}{%
 \providecommand\@@startlink[1]{%
  \leavevmode
  \pdfstartlink
   attr{/Border[0 0 1 ]/H/I/C[0 1 1]}%
   user{/Subtype/Link/A<</Type/Action/S/URI/URI(#1)>>}%
  \relax
 }%
 \providecommand\@@endlink[0]{\pdfendlink}%
}%
\providecommand \url  [0]{\begingroup\@sanitize \@url }%
\providecommand \@url [1]{\endgroup\@href {#1}{\urlprefix}}%
\providecommand \urlprefix [0]{URL }%
\providecommand \Eprint[0]{\href }%
\@ifxundefined \urlstyle {%
  \providecommand \doi [1]{doi:\discretionary{}{}{}#1}%
}{%
  \providecommand \doi [0]{doi:\discretionary{}{}{}\begingroup
  \urlstyle{rm}\Url }%
}%
\providecommand \doibase [0]{http://dx.doi.org/}%
\providecommand \Doi[1]{\href{\doibase#1}}%
\providecommand \bibAnnote [3]{%
  \BibitemShut{#1}%
  \begin{quotation}\noindent
    \textsc{Key:}\ #2\\\textsc{Annotation:}\ #3%
  \end{quotation}%
}%
\providecommand \bibAnnoteFile [2]{%
  \IfFileExists{#2}{\bibAnnote {#1} {#2} {\input{#2}}}{}%
}%
\providecommand \typeout [0]{\immediate \write \m@ne }%
\providecommand \selectlanguage [0]{\@gobble}%
\providecommand \bibinfo [0]{\@secondoftwo}%
\providecommand \bibfield [0]{\@secondoftwo}%
\providecommand \translation [1]{[#1]}%
\providecommand \BibitemOpen[0]{}%
\providecommand \bibitemStop [0]{}%
\providecommand \bibitemNoStop [0]{.\EOS\space}%
\providecommand \EOS [0]{\spacefactor3000\relax}%
\providecommand \BibitemShut [1]{\csname bibitem#1\endcsname}%
\bibitem{Mehrer2007book}%
  \BibitemOpen
  \bibfield{author}{%
  \bibinfo {author} {\bibfnamefont{H.}~\bibnamefont{Mehrer}},\ }%
  \enquote{\bibinfo {title} {Diffusion in solids},}\ \ (\bibinfo {publisher}
  {Springer-Verlag},\ \bibinfo {year} {2007})%
  \bibAnnoteFile{NoStop}{Mehrer2007book}%
\bibitem{Tan1991a}%
  \BibitemOpen
  \bibfield{author}{%
  \bibinfo {author} {\bibfnamefont{T.~Y.}\ \bibnamefont{Tan}}, \bibinfo
  {author} {\bibfnamefont{U.}~\bibnamefont{G{\"o}sele}},\ and\ \bibinfo
  {author} {\bibfnamefont{S.}~\bibnamefont{Yu}},\ }%
  \bibfield{journal}{%
  \bibinfo {journal} {Crit.\ Rev.\ in Sol.\ State and Mater.\ Sci.}\ }%
  \textbf{\bibinfo {volume} {17}},\ \bibinfo {pages} {47} (\bibinfo {year}
  {1991})%
  \bibAnnoteFile{NoStop}{Tan1991a}%
\bibitem{Bracht2005a}%
  \BibitemOpen
  \bibfield{author}{%
  \bibinfo {author} {\bibfnamefont{H.}~\bibnamefont{Bracht}}\ and\ \bibinfo
  {author} {\bibfnamefont{S.}~\bibnamefont{Brotzmann}},\ }%
  \bibfield{journal}{%
  \Doi{10.1103/PhysRevB.71.115216}{\bibinfo {journal} {Phys.\ Rev.\ B}}\ }%
  \textbf{\bibinfo {volume} {71}},\ \bibinfo {pages} {115216} (\bibinfo {year} {2005})%
  \bibAnnoteFile{NoStop}{Bracht2005a}%
\bibitem{Koumetz2006a}%
  \BibitemOpen
  \bibfield{author}{%
  \bibinfo {author} {\bibfnamefont{S.}~\bibnamefont{Koumetz}}, \bibinfo
  {author} {\bibfnamefont{J.~C.}\ \bibnamefont{Pesant}},\ and\ \bibinfo
  {author} {\bibfnamefont{C.}~\bibnamefont{Dubois}},\ }%
  \bibfield{journal}{%
  \bibinfo {journal} {J.\ Phys.\ Condens.\ Matter}\ }%
  \textbf{\bibinfo {volume} {18}},\ \bibinfo {pages} {L283} (\bibinfo {year}
  {2006})%
  \bibAnnoteFile{NoStop}{Koumetz2006a}%
\bibitem{Zucker1989a}%
  \BibitemOpen
  \bibfield{author}{%
  \bibinfo {author} {\bibfnamefont{E.~P.}\ \bibnamefont{Zucker}}, \bibinfo
  {author} {\bibfnamefont{A.}~\bibnamefont{Hashimoto}}, \bibinfo {author}
  {\bibfnamefont{T.}~\bibnamefont{Fukunaga}},\ and\ \bibinfo {author}
  {\bibfnamefont{N.}~\bibnamefont{Watanabe}},\ }%
  \bibfield{journal}{%
  \Doi{10.1063/1.100932}{\bibinfo {journal} {Appl.\ Phys.\ Lett.}}\ }%
  \textbf{\bibinfo {volume} {54}},\ \bibinfo {pages} {564} (\bibinfo {year}
  {1989})%
  \bibAnnoteFile{NoStop}{Zucker1989a}%
\bibitem{Bosker1995a}%
  \BibitemOpen
  \bibfield{author}{%
  \bibinfo {author} {\bibfnamefont{G.}~\bibnamefont{{B\"osker}}}, \bibinfo
  {author} {\bibfnamefont{N.~A.}\ \bibnamefont{Stolwijk}}, \bibinfo {author}
  {\bibfnamefont{H.-G.}\ \bibnamefont{Hettwer}}, \bibinfo {author}
  {\bibfnamefont{A.}~\bibnamefont{Rucki}}, \bibinfo {author}
  {\bibfnamefont{W.}~\bibnamefont{{J\"ager}}},\ and\ \bibinfo {author}
  {\bibfnamefont{U.}~\bibnamefont{{S\"odervall}}},\ }%
  \bibfield{journal}{%
  \Doi{10.1103/PhysRevB.52.11927}{\bibinfo {journal} {Phys.\ Rev.\ B}}\ }%
  \textbf{\bibinfo {volume} {52}},\ \bibinfo {pages} {11927} (\bibinfo {year} {1995})%
  \bibAnnoteFile{NoStop}{Bosker1995a}%
\bibitem{Bosker1999a}%
  \BibitemOpen
  \bibfield{author}{%
  \bibinfo {author} {\bibfnamefont{G.}~\bibnamefont{B\"{o}sker}}, \bibinfo
  {author} {\bibfnamefont{N.~A.}\ \bibnamefont{Stolwijk}}, \bibinfo {author}
  {\bibfnamefont{H.}~\bibnamefont{Mehrer}}, \bibinfo {author}
  {\bibfnamefont{U.}~\bibnamefont{S\"{o}dervall}},\ and\ \bibinfo {author}
  {\bibfnamefont{W.}~\bibnamefont{J\"{a}ger}},\ }%
  \bibfield{journal}{%
  \Doi{10.1063/1.370806}{\bibinfo {journal} {Journal of Applied Physics}}\ }%
  \textbf{\bibinfo {volume} {86}},\ \bibinfo {pages} {791} (\bibinfo {year}
  {1999})%
  \bibAnnoteFile{NoStop}{Bosker1999a}%
\bibitem{Bracht2001a}%
  \BibitemOpen
  \bibfield{author}{%
  \bibinfo {author} {\bibfnamefont{H.}~\bibnamefont{Bracht}}, \bibinfo {author}
  {\bibfnamefont{M.~S.}\ \bibnamefont{Norseng}}, \bibinfo {author}
  {\bibfnamefont{E.~E.}\ \bibnamefont{Haller}},\ and\ \bibinfo {author}
  {\bibfnamefont{K.}~\bibnamefont{Eberl}},\ }%
  \bibfield{journal}{%
  \Doi{DOI: 10.1016/S0921-4526(01)00817-1}{\bibinfo {journal} {Physica B}}\ }%
  \textbf{\bibinfo {volume} {308-310}},\ \bibinfo {pages} {831} (\bibinfo
  {year} {2001}),\ ISSN \bibinfo {issn} {0921-4526}%
  \bibAnnoteFile{NoStop}{Bracht2001a}%
\bibitem{BrachtPrivate}%
  \BibitemOpen
  \bibfield{author}{%
  \bibinfo {author} {\bibfnamefont{H.}~\bibnamefont{Bracht}},\ }%
  \bibinfo {note} {private communication}%
  \bibAnnoteFile{NoStop}{BrachtPrivate}%
\bibitem{Zhang1991}%
  \BibitemOpen
  \bibfield{author}{%
  \bibinfo {author} {\bibfnamefont{S.~B.}\ \bibnamefont{Zhang}}\ and\ \bibinfo
  {author} {\bibfnamefont{J.~E.}\ \bibnamefont{Northrup}},\ }%
  \bibfield{journal}{%
  \bibinfo {journal} {Phys.\ Rev.\ Lett.}\ }%
  \textbf{\bibinfo {volume} {67}},\ \bibinfo {pages} {2339} (\bibinfo {year}
  {1991})%
  \bibAnnoteFile{NoStop}{Zhang1991}%
\bibitem{Chadi1992b}%
  \BibitemOpen
  \bibfield{author}{%
  \bibinfo {author} {\bibfnamefont{D.~J.}\ \bibnamefont{Chadi}},\ }%
  \bibfield{journal}{%
  \bibinfo {journal} {Phys.\ Rev.\ B}\ }%
  \textbf{\bibinfo {volume} {46}},\ \bibinfo {pages} {15053} (\bibinfo {year}
  {1992})%
  \bibAnnoteFile{NoStop}{Chadi1992b}%
\bibitem{Chadi1992}%
  \BibitemOpen
  \bibfield{author}{%
  \bibinfo {author} {\bibfnamefont{D.~J.}\ \bibnamefont{Chadi}},\ }%
  \bibfield{journal}{%
  \Doi{10.1103/PhysRevB.46.9400}{\bibinfo {journal} {Phys. Rev. B}}\ }%
  \textbf{\bibinfo {volume} {46}},\ \bibinfo {pages} {9400} (\bibinfo {year} {1992})%
  \bibAnnoteFile{NoStop}{Chadi1992}%
\bibitem{Landman1997}%
  \BibitemOpen
  \bibfield{author}{%
  \bibinfo {author} {\bibfnamefont{J.~I.}\ \bibnamefont{Landman}}, \bibinfo
  {author} {\bibfnamefont{C.~G.}\ \bibnamefont{Morgan}}, \bibinfo {author}
  {\bibfnamefont{J.~T.}\ \bibnamefont{Schick}}, \bibinfo {author}
  {\bibfnamefont{P.}~\bibnamefont{Papoulias}},\ and\ \bibinfo {author}
  {\bibfnamefont{A.}~\bibnamefont{Kumar}},\ }%
  \bibfield{journal}{%
  \Doi{10.1103/PhysRevB.55.15581}{\bibinfo {journal} {Phys.\ Rev.\ B}}\ }%
  \textbf{\bibinfo {volume} {55}},\ \bibinfo {pages} {15581} (\bibinfo {year} {1997})%
  \bibAnnoteFile{NoStop}{Landman1997}%
\bibitem{Schick2002a}%
  \BibitemOpen
  \bibfield{author}{%
  \bibinfo {author} {\bibfnamefont{J.~T.}\ \bibnamefont{Schick}}, \bibinfo
  {author} {\bibfnamefont{C.~G.}\ \bibnamefont{Morgan}},\ and\ \bibinfo
  {author} {\bibfnamefont{P.}~\bibnamefont{Papoulias}},\ }%
  \bibfield{journal}{%
  \Doi{10.1103/PhysRevB.66.195302}{\bibinfo {journal} {Phys.\ Rev.\ B}}\ }%
  \textbf{\bibinfo {volume} {66}},\ \bibinfo {pages} {195302} (\bibinfo {year} {2002})%
  \bibAnnoteFile{NoStop}{Schick2002a}%
\bibitem{Bar-Yam1984a}%
  \BibitemOpen
  \bibfield{author}{%
  \bibinfo {author} {\bibfnamefont{Y.}~\bibnamefont{Bar-Yam}}\ and\ \bibinfo
  {author} {\bibfnamefont{J.~D.}\ \bibnamefont{Joannopoulos}},\ }%
  \bibfield{journal}{%
  \Doi{10.1103/PhysRevB.30.1844}{\bibinfo {journal} {Phys. Rev. B}}\ }%
  \textbf{\bibinfo {volume} {30}},\ \bibinfo {pages} {1844} (\bibinfo {year} {1984})%
  \bibAnnoteFile{NoStop}{Bar-Yam1984a}%
\bibitem{Pantelides1983a}%
  \BibitemOpen
  \bibfield{author}{%
  \bibinfo {author} {\bibfnamefont{S.}~\bibnamefont{Pantelides}}, \bibinfo
  {author} {\bibfnamefont{I.}~\bibnamefont{Ivanov}}, \bibinfo {author}
  {\bibfnamefont{M.}~\bibnamefont{Scheffler}},\ and\ \bibinfo {author}
  {\bibfnamefont{J.}~\bibnamefont{Vigneron}},\ }%
  \bibfield{journal}{%
  \bibinfo {journal} {Physica B \& C}\ }%
  \textbf{\bibinfo {volume} {116}},\ \bibinfo {pages} {18} (\bibinfo {year}
  {1983}),\ ISSN \bibinfo {issn} {0378-4371}%
  \bibAnnoteFile{NoStop}{Pantelides1983a}%
\bibitem{Needs1999a}%
  \BibitemOpen
  \bibfield{author}{%
  \bibinfo {author} {\bibfnamefont{R.~J.}\ \bibnamefont{Needs}},\ }%
  \bibfield{journal}{%
  \bibinfo {journal} {Journal of Physics: Condensed Matter}\ }%
  \textbf{\bibinfo {volume} {11}},\ \bibinfo {pages} {10437} (\bibinfo {year}
  {1999})%
\bibitem{Goedecker2002a}%
  \BibitemOpen
  \bibfield{author}{%
  \bibinfo {author} {\bibfnamefont{S.}~\bibnamefont{Goedecker}}, \bibinfo
  {author} {\bibfnamefont{T.}~\bibnamefont{Deutsch}},\ and\ \bibinfo {author}
  {\bibfnamefont{L.}~\bibnamefont{Billard}},\ }%
  \bibfield{journal}{%
  \Doi{10.1103/PhysRevLett.88.235501}{\bibinfo {journal} {Phys. Rev. Lett.}}\
  }%
  \textbf{\bibinfo {volume} {88}},\ \bibinfo {pages} {235501} (\bibinfo {month}
  {May}\ \bibinfo {year} {2002})%
  \bibAnnoteFile{NoStop}{Goedecker2002a}%
\bibitem{Puska1998}%
  \BibitemOpen
  \bibfield{author}{%
  \bibinfo {author} {\bibfnamefont{M.~J.}\ \bibnamefont{Puska}}, \bibinfo
  {author} {\bibfnamefont{S.}~\bibnamefont{{P\"oykk\"o}}}, \bibinfo {author}
  {\bibfnamefont{M.}~\bibnamefont{Pesola}},\ and\ \bibinfo {author}
  {\bibfnamefont{R.~M.}\ \bibnamefont{Nieminen}},\ }%
  \bibfield{journal}{%
  \bibinfo {journal} {Phys.\ Rev.\ B}\ }%
  \textbf{\bibinfo {volume} {58}},\ \bibinfo {pages} {1318} (\bibinfo {year}
  {1998})%
  \bibAnnoteFile{NoStop}{Puska1998}%
\bibitem{Schultz2009a}%
  \BibitemOpen
  \bibfield{author}{%
  \bibinfo {author} {\bibfnamefont{P.~A.}\ \bibnamefont{Schultz}}\ and\
  \bibinfo {author} {\bibfnamefont{O.~A.}\ \bibnamefont{von Lilienfeld}},\ }%
  \bibfield{journal}{%
  \bibinfo {journal} {Modelling and Simulation in Materials Science and
  Engineering}\ }%
  \textbf{\bibinfo {volume} {17}},\ \bibinfo {pages} {084007} (\bibinfo {year}
  {2009})%
  \bibAnnoteFile{NoStop}{Schultz2009a}%
\bibitem{Malouin2007a}%
  \BibitemOpen
  \bibfield{author}{%
  \bibinfo {author} {\bibfnamefont{M.-A.}\ \bibnamefont{Malouin}}, \bibinfo
  {author} {\bibfnamefont{F.}~\bibnamefont{El-Mallouhi}},\ and\ \bibinfo
  {author} {\bibfnamefont{N.}~\bibnamefont{Mousseau}},\ }%
  \bibfield{journal}{%
  \bibinfo {journal} {Phys.\ Rev.\ B}\ }%
  \textbf{\bibinfo {volume} {76}},\ \bibinfo {pages} {045211} (\bibinfo {year}
  {2007})%
  \bibAnnoteFile{NoStop}{Malouin2007a}%
\bibitem{Bockstedte1997b}%
  \BibitemOpen
  \bibfield{author}{%
  \bibinfo {author} {\bibfnamefont{M.}~\bibnamefont{Bockstedte}}\ and\ \bibinfo
  {author} {\bibfnamefont{M.}~\bibnamefont{Scheffler}},\ }%
  \bibfield{journal}{%
  \bibinfo {journal} {Z.\ Phys.\ Chem. (Munich)}\ }%
  \textbf{\bibinfo {volume} {200}},\ \bibinfo {pages} {195} (\bibinfo {year}
  {1997})%
  \bibAnnoteFile{NoStop}{Bockstedte1997b}%
\bibitem{ElMellouhi2005a}%
  \BibitemOpen
  \bibfield{author}{%
  \bibinfo {author} {\bibfnamefont{F.}~\bibnamefont{El-Mellouhi}}\ and\
  \bibinfo {author} {\bibfnamefont{N.}~\bibnamefont{Mousseau}},\ }%
  \bibfield{journal}{%
  \bibinfo {journal} {Phys.\ Rev.\ B}\ }%
  \textbf{\bibinfo {volume} {71}},\ \bibinfo {pages} {125207} (\bibinfo {year}
  {2005})%
  \bibAnnoteFile{NoStop}{ElMellouhi2005a}%
\bibitem{ElMellouhi2006a}%
  \BibitemOpen
  \bibfield{author}{%
  \bibinfo {author} {\bibfnamefont{F.}~\bibnamefont{El-Mellouhi}}\ and\
  \bibinfo {author} {\bibfnamefont{N.}~\bibnamefont{Mousseau}},\ }%
  \bibfield{journal}{%
  \bibinfo {journal} {Phys.\ Rev.\ B}\ }%
  \textbf{\bibinfo {volume} {74}},\ \bibinfo {pages} {205207} (\bibinfo {year}
  {2006})%
  \bibAnnoteFile{NoStop}{ElMellouhi2006a}%
\bibitem{Papoulias2010}%
  \BibitemOpen
  \bibfield{author}{%
  \bibinfo {author} {\bibfnamefont{P.~G.}\ \bibnamefont{Papoulias}}, \bibinfo
  {author} {\bibfnamefont{C.~G.}\ \bibnamefont{Morgan}},\ and\ \bibinfo
  {author} {\bibfnamefont{J.~T.}\ \bibnamefont{Schick}},\ }%
  \bibfield{journal}{%
  \bibinfo {journal} {submitted}}%
   (\bibinfo {year} {2011})%
  \bibAnnoteFile{NoStop}{Papoulias2010}%
\bibitem{Papoulias2009}%
  \BibitemOpen
  \bibfield{author}{%
  \bibinfo {author} {\bibfnamefont{P.~G.}\ \bibnamefont{Papoulias}},\ }%
  Ph.D. thesis,\ \bibinfo {school} {Wayne State University} (\bibinfo {year}
  {2009})%
  \bibAnnoteFile{NoStop}{Papoulias2009}%
\bibitem{Levasseur-Smith2008a}%
  \BibitemOpen
  \bibfield{author}{%
  \bibinfo {author} {\bibfnamefont{K.}~\bibnamefont{Levasseur-Smith}}\ and\
  \bibinfo {author} {\bibfnamefont{N.}~\bibnamefont{Mousseau}},\ }%
  \bibfield{journal}{%
  \bibinfo {journal} {J. Appl.\ Phys.}\ }%
  \textbf{\bibinfo {volume} {103}},\ \bibinfo {pages} {113502} (\bibinfo {year}
  {2008})%
  \bibAnnoteFile{NoStop}{Levasseur-Smith2008a}%
\bibitem{Levasseur-Smith2008b}%
  \BibitemOpen
  \bibfield{author}{%
  \bibinfo {author} {\bibfnamefont{K.}~\bibnamefont{Levasseur-Smith}}\ and\
  \bibinfo {author} {\bibfnamefont{N.}~\bibnamefont{Mousseau}},\ }%
  \bibfield{journal}{%
  \bibinfo {journal} {Eur.\ Phys.\ J.\ B}\ }%
  \textbf{\bibinfo {volume} {64}},\ \bibinfo {pages} {165} (\bibinfo {year}
  {2008})%
  \bibAnnoteFile{NoStop}{Levasseur-Smith2008b}%
\bibitem{Bockstedte1997}%
  \BibitemOpen
  \bibfield{author}{%
  \bibinfo {author} {\bibfnamefont{M.}~\bibnamefont{Bockstedte}}, \bibinfo
  {author} {\bibfnamefont{A.}~\bibnamefont{Kley}}, \bibinfo {author}
  {\bibfnamefont{J.}~\bibnamefont{Neugebauer}},\ and\ \bibinfo {author}
  {\bibfnamefont{M.}~\bibnamefont{Scheffler}},\ }%
  \bibfield{journal}{%
  \bibinfo {journal} {Comput.\ Phys.\ Commun.}\ }%
  \textbf{\bibinfo {volume} {107}},\ \bibinfo {pages} {187} (\bibinfo {year}
  {1997})%
  \bibAnnoteFile{NoStop}{Bockstedte1997}%
\bibitem{vasp1}%
  \BibitemOpen
  \bibfield{author}{%
  \bibinfo {author} {\bibfnamefont{G.}~\bibnamefont{Kresse}}\ and\ \bibinfo
  {author} {\bibfnamefont{J.}~\bibnamefont{Hafner}},\ }%
  \bibfield{journal}{%
  \Doi{10.1103/PhysRevB.47.558}{\bibinfo {journal} {Phys. Rev. B}}\ }%
  \textbf{\bibinfo {volume} {47}},\ \bibinfo {pages} {558} (\bibinfo {month}
  {Jan}\ \bibinfo {year} {1993})%
  \bibAnnoteFile{NoStop}{vasp1}%
\bibitem{vasp2}%
  \BibitemOpen
  \bibfield{author}{%
  \bibinfo {author} {\bibfnamefont{G.}~\bibnamefont{Kresse}},\ }%
  Ph.D. thesis,\ \bibinfo {school} {Technische Universit\"at Wien} (\bibinfo
  {year} {1993})%
  \bibAnnoteFile{NoStop}{vasp2}%
\bibitem{vasp3}%
  \BibitemOpen
  \bibfield{author}{%
  \bibinfo {author} {\bibfnamefont{J.~F.}\ \bibnamefont{G.~Kresse}},\ }%
  \bibfield{journal}{%
  \bibinfo {journal} {Comput. Mat. Sci.}\ }%
  \textbf{\bibinfo {volume} {6}},\ \bibinfo {pages} {15} (\bibinfo {year} {1996})%
  \bibAnnoteFile{NoStop}{vasp3}%
\bibitem{vasp4}%
  \BibitemOpen
  \bibfield{author}{%
  \bibinfo {author} {\bibfnamefont{G.}~\bibnamefont{Kresse}}\ and\ \bibinfo
  {author} {\bibfnamefont{J.}~\bibnamefont{Furthm\"uller}},\ }%
  \bibfield{journal}{%
  \Doi{10.1103/PhysRevB.54.11169}{\bibinfo {journal} {Phys. Rev. B}}\ }%
  \textbf{\bibinfo {volume} {54}},\ \bibinfo {pages} {11169} (\bibinfo {month}
  {Oct}\ \bibinfo {year} {1996})%
  \bibAnnoteFile{NoStop}{vasp4}%
\bibitem{Hohenberg1964}%
  \BibitemOpen
  \bibfield{author}{%
  \bibinfo {author} {\bibfnamefont{P.}~\bibnamefont{Hohenberg}}\ and\ \bibinfo
  {author} {\bibfnamefont{W.}~\bibnamefont{Kohn}},\ }%
  \bibfield{journal}{%
  \bibinfo {journal} {Phys.\ Rev.}\ }%
  \textbf{\bibinfo {volume} {136}},\ \bibinfo {pages} {B864} (\bibinfo {year}
  {1964})%
  \bibAnnoteFile{NoStop}{Hohenberg1964}%
\bibitem{Ceperley1980}%
  \BibitemOpen
  \bibfield{author}{%
  \bibinfo {author} {\bibfnamefont{D.~M.}\ \bibnamefont{Ceperley}}\ and\
  \bibinfo {author} {\bibfnamefont{G.~J.}\ \bibnamefont{Alder}},\ }%
  \bibfield{journal}{%
  \bibinfo {journal} {Phys.\ Rev.\ Lett.}\ }%
  \textbf{\bibinfo {volume} {45}},\ \bibinfo {pages} {566} (\bibinfo {year}
  {1980})%
  \bibAnnoteFile{NoStop}{Ceperley1980}%
\bibitem{Perdew1981}%
  \BibitemOpen
  \bibfield{author}{%
  \bibinfo {author} {\bibfnamefont{J.}~\bibnamefont{Perdew}}\ and\ \bibinfo
  {author} {\bibfnamefont{A.}~\bibnamefont{Zunger}},\ }%
  \bibfield{journal}{%
  \bibinfo {journal} {Phys.\ Rev.\ B}\ }%
  \textbf{\bibinfo {volume} {23}},\ \bibinfo {pages} {5048} (\bibinfo {year}
  {1981})%
  \bibAnnoteFile{NoStop}{Perdew1981}%
\bibitem{Kleinman1982}%
  \BibitemOpen
  \bibfield{author}{%
  \bibinfo {author} {\bibfnamefont{L.}~\bibnamefont{Kleinman}}\ and\ \bibinfo
  {author} {\bibfnamefont{D.~M.}\ \bibnamefont{Bylander}},\ }%
  \bibfield{journal}{%
  \bibinfo {journal} {Phys.\ Rev.\ Lett.}\ }%
  \textbf{\bibinfo {volume} {48}},\ \bibinfo {pages} {1425} (\bibinfo {year}
  {1982})%
  \bibAnnoteFile{NoStop}{Kleinman1982}%
\bibitem{Hamann1989}%
  \BibitemOpen
  \bibfield{author}{%
  \bibinfo {author} {\bibfnamefont{D.~R.}\ \bibnamefont{Hamann}},\ }%
  \bibfield{journal}{%
  \bibinfo {journal} {Phys.\ Rev.\ B}\ }%
  \textbf{\bibinfo {volume} {40}},\ \bibinfo {pages} {2980} (\bibinfo {year}
  {1989})%
  \bibAnnoteFile{NoStop}{Hamann1989}%
\bibitem{vanderbilt1}%
  \BibitemOpen
  \bibfield{author}{%
  \bibinfo {author} {\bibfnamefont{D.}~\bibnamefont{Vanderbilt}},\ }%
  \bibfield{journal}{%
  \Doi{10.1103/PhysRevB.41.7892}{\bibinfo {journal} {Phys. Rev. B}}\ }%
  \textbf{\bibinfo {volume} {41}},\ \bibinfo {pages} {7892} (\bibinfo {month}
  {Apr}\ \bibinfo {year} {1990})%
  \bibAnnoteFile{NoStop}{vanderbilt1}%
\bibitem{vanderbilt2}%
  \BibitemOpen
  \bibfield{author}{%
  \bibinfo {author} {\bibfnamefont{G.}~\bibnamefont{Kresse}}\ and\ \bibinfo
  {author} {\bibfnamefont{J.}~\bibnamefont{Hafner}},\ }%
  \bibfield{journal}{%
  \bibinfo {journal} {J.\ Phys.\ Condens.\ Mat.}\ }%
  \textbf{\bibinfo {volume} {6}},\ \bibinfo {pages} {8245} (\bibinfo {year}
  {1994})%
  \bibAnnoteFile{NoStop}{vanderbilt2}%
\bibitem{Jonsson1998a}%
  \BibitemOpen
  \bibfield{author}{%
  \bibinfo {author} {\bibfnamefont{H.}~\bibnamefont{J{\'o}nsson}}, \bibinfo
  {author} {\bibfnamefont{G.}~\bibnamefont{Mills}},\ and\ \bibinfo {author}
  {\bibfnamefont{K.~W.}\ \bibnamefont{Jacobsen}},\ }%
  \enquote{\bibinfo {title} {Classical and quantum dynamics in condensed phase
  systems},}\ \ (\bibinfo {publisher} {World Scientific},\ \bibinfo {year}
  {1998})\ Chap.~\bibinfo {chapter} {16}, pp.\ \bibinfo {pages} {385--404}%
  \bibAnnoteFile{NoStop}{Jonsson1998a}%
\bibitem{Mills1995a}%
  \BibitemOpen
  \bibfield{author}{%
  \bibinfo {author} {\bibfnamefont{G.}~\bibnamefont{Mills}}, \bibinfo {author}
  {\bibfnamefont{H.}~\bibnamefont{J{\'o}nsson}},\ and\ \bibinfo {author}
  {\bibfnamefont{G.~K.}\ \bibnamefont{Schenter}},\ }%
  \bibfield{journal}{%
  \bibinfo {journal} {Surf. Sci.}\ }%
  \textbf{\bibinfo {volume} {324}},\ \bibinfo {pages} {305} (\bibinfo {year} {1995})%
  \bibAnnoteFile{NoStop}{Mills1995a}%
\bibitem{Murnaghan1944a}%
  \BibitemOpen
  \bibfield{author}{%
  \bibinfo {author} {\bibfnamefont{F.~D.}\ \bibnamefont{Murnaghan}},\ }%
  \bibfield{journal}{%
  \bibinfo {journal} {Proc.\ Natl.\ Acad.\ Sci.}\ }%
  \textbf{\bibinfo {volume} {30}},\ \bibinfo {pages} {244} (\bibinfo {year}
  {1944})%
  \bibAnnoteFile{NoStop}{Murnaghan1944a}%
\bibitem{Blakemore1982a}%
  \BibitemOpen
  \bibfield{author}{%
  \bibinfo {author} {\bibfnamefont{J.~S.}\ \bibnamefont{Blakemore}},\ }%
  \bibfield{journal}{%
  \Doi{10.1063/1.331665}{\bibinfo {journal} {J.\ Appl.\ Phys.}}\ }%
  \textbf{\bibinfo {volume} {53}},\ \bibinfo {pages} {R123} (\bibinfo {year}
  {1982})%
  \bibAnnoteFile{NoStop}{Blakemore1982a}%
\bibitem{VESTA}%
  \BibitemOpen
  \bibfield{author}{%
  \bibinfo {author} {\bibfnamefont{K.}~\bibnamefont{Momma}}\ and\ \bibinfo
  {author} {\bibfnamefont{F.}~\bibnamefont{Izumi}},\ }%
  \bibfield{journal}{%
  \Doi{10.1107/S0021889808012016}{\bibinfo {journal} {J.\ Appl.\ Crystallog.}}\ }%
  \textbf{\bibinfo {volume} {41}},\ \bibinfo {pages} {653} (\bibinfo {year} {2008})%
  \bibAnnoteFile{NoStop}{VESTA}%
\bibitem{Kohan2000}%
  \BibitemOpen
  \bibfield{author}{%
  \bibinfo {author} {\bibfnamefont{A.~F.}\ \bibnamefont{Kohan}}, \bibinfo
  {author} {\bibfnamefont{G.}~\bibnamefont{Ceder}}, \bibinfo {author}
  {\bibfnamefont{D.}~\bibnamefont{Morgan}},\ and\ \bibinfo {author}
  {\bibfnamefont{C.~G.}\ \bibnamefont{{Van de Walle}}},\ }%
  \bibfield{journal}{%
  \bibinfo {journal} {Phys.\ Rev.\ B}\ }%
  \textbf{\bibinfo {volume} {61}},\ \bibinfo {pages} {15019} (\bibinfo {year}
  {2000})%
  \bibAnnoteFile{NoStop}{Kohan2000}%
\bibitem{Makov1995a}%
  \BibitemOpen
  \bibfield{author}{%
  \bibinfo {author} {\bibfnamefont{G.}~\bibnamefont{Makov}}\ and\ \bibinfo
  {author} {\bibfnamefont{M.~C.}\ \bibnamefont{Payne}},\ }%
  \bibfield{journal}{%
  \bibinfo {journal} {Phys.\ Rev.\ B}\ }%
  \textbf{\bibinfo {volume} {51}},\ \bibinfo {pages} {4014} (\bibinfo {year}
  {1995})%
  \bibAnnoteFile{NoStop}{Makov1995a}%
\bibitem{Vandewalle2004a}%
  \BibitemOpen
  \bibfield{author}{%
  \bibinfo {author} {\bibfnamefont{C.~G.}\ \bibnamefont{{Van de Walle}}}\ and\
  \bibinfo {author} {\bibfnamefont{J.}~\bibnamefont{Neugebauer}},\ }%
  \bibfield{journal}{%
  \bibinfo {journal} {J. Appl.\ Phys.}\ }%
  \textbf{\bibinfo {volume} {95}},\ \bibinfo {pages} {3851} (\bibinfo {month}
  {April}\ \bibinfo {year} {2004})%
  \bibAnnoteFile{NoStop}{Vandewalle2004a}%
\bibitem{freysoldt2009a}%
  \BibitemOpen
  \bibfield{author}{%
  \bibinfo {author} {\bibfnamefont{C.}~\bibnamefont{Freysoldt}}, \bibinfo
  {author} {\bibfnamefont{J.}~\bibnamefont{Neugebauer}},\ and\ \bibinfo
  {author} {\bibfnamefont{C.~G.}\ \bibnamefont{{Van de Walle}}},\ }%
  \bibfield{journal}{%
  \Doi{10.1103/PhysRevLett.102.016402}{\bibinfo {journal} {Phys.\ Rev.\ Lett.}}\ }%
  \textbf{\bibinfo {volume} {102}},\ \bibinfo {eid} {016402} (\bibinfo {year}
  {2009})%
  \bibAnnoteFile{NoStop}{freysoldt2009a}%
\bibitem{Niemenen2007a}%
  \BibitemOpen
  \bibfield{author}{%
  \bibinfo {author} {\bibfnamefont{R.~M.}\ \bibnamefont{Nieminen}},\ }%
  \bibfield{journal}{%
  \bibinfo {journal} {Topics Appl.\ Physics}\ }%
  \textbf{\bibinfo {volume} {104}},\ \bibinfo {pages} {29} (\bibinfo {year}
  {2007})%
  \bibAnnoteFile{NoStop}{Niemenen2007a}%
\bibitem{Baraff1985b}%
  \BibitemOpen
  \bibfield{author}{%
  \bibinfo {author} {\bibfnamefont{G.~A.}\ \bibnamefont{Baraff}}\ and\ \bibinfo
  {author} {\bibfnamefont{M.}~\bibnamefont{{Schl\"uter}}},\ }%
  \bibfield{journal}{%
  \bibinfo {journal} {Phys.\ Rev.\ Lett.}\ }%
  \textbf{\bibinfo {volume} {55}},\ \bibinfo {pages} {2340} (\bibinfo {year}
  {1985})%
  \bibAnnoteFile{NoStop}{Baraff1985b}%
\bibitem{Baraff1986}%
  \BibitemOpen
  \bibfield{author}{%
  \bibinfo {author} {\bibfnamefont{G.~A.}\ \bibnamefont{Baraff}}\ and\ \bibinfo
  {author} {\bibfnamefont{M.}~\bibnamefont{{Schl\"uter}}},\ }%
  \bibfield{journal}{%
  \bibinfo {journal} {Phys.\ Rev.\ B}\ }%
  \textbf{\bibinfo {volume} {33}},\ \bibinfo {pages} {7346} (\bibinfo {year}
  {1986})%
  \bibAnnoteFile{NoStop}{Baraff1986}%
\bibitem{Note1}%
  \BibitemOpen
  \bibinfo {note} {Differences in the formation energies presented here and our
  previously published calculations\cite {Schick2002a} are a result of having
  now evaluated the bulk arsenic structure within our own computations, rather
  than relying upon a previous calculation that used the same codes. This
  brings our formation energies into closer agreement with other work. For
  example, the neutral arsenic antisite formation energy in the arsenic-rich
  limit was stated to be 1.8~eV in our earlier work, and with the updated bulk
  arsenic formation energy it is now 1.3~eV, which agrees with the calculation
  due to Schultz {\protect \em et al.}\cite {Schultz2009a} to within the
  expected precision of density functional theory.}%
  \bibAnnoteFile{Stop}{Note1}%
\bibitem{Lany2008a}%
  \BibitemOpen
  \bibfield{author}{%
  \bibinfo {author} {\bibfnamefont{S.}~\bibnamefont{Lany}}\ and\ \bibinfo
  {author} {\bibfnamefont{A.}~\bibnamefont{Zunger}},\ }%
  \bibfield{journal}{%
  \Doi{10.1103/PhysRevB.78.235104}{\bibinfo {journal} {Phys. Rev. B}}\ }%
  \textbf{\bibinfo {volume} {78}},\ \bibinfo {pages} {235104} (\bibinfo {year} {2008})%
  \bibAnnoteFile{NoStop}{Lany2008a}%
\bibitem{Schultz2006a}%
  \BibitemOpen
  \bibfield{author}{%
  \bibinfo {author} {\bibfnamefont{P.~A.}\ \bibnamefont{Schultz}},\ }%
  \bibfield{journal}{%
  \Doi{10.1103/PhysRevLett.96.246401}{\bibinfo {journal} {Phys. Rev. Lett.}}\
  }%
  \textbf{\bibinfo {volume} {96}},\ \bibinfo {pages} {246401} (\bibinfo {year} {2006})%
  \bibAnnoteFile{NoStop}{Schultz2006a}%
\bibitem{Tuttle2008a}%
  \BibitemOpen
  \bibfield{author}{%
  \bibinfo {author} {\bibfnamefont{B.~R.}\ \bibnamefont{Tuttle}}\ and\ \bibinfo
  {author} {\bibfnamefont{S.~T.}\ \bibnamefont{Pantelides}},\ }%
  \bibfield{journal}{%
  \Doi{10.1103/PhysRevLett.101.089701}{\bibinfo {journal} {Phys. Rev. Lett.}}\
  }%
  \textbf{\bibinfo {volume} {101}},\ \bibinfo {pages} {089701} (\bibinfo {year} {2008})%
  \bibAnnoteFile{NoStop}{Tuttle2008a}%
\bibitem{Schultz2008a}%
  \BibitemOpen
  \bibfield{author}{%
  \bibinfo {author} {\bibfnamefont{P.~A.}\ \bibnamefont{Schultz}},\ }%
  \bibfield{journal}{%
  \Doi{10.1103/PhysRevLett.101.089702}{\bibinfo {journal} {Phys.\ Rev.\ Lett.}}\
  }%
  \textbf{\bibinfo {volume} {101}},\ \bibinfo {pages} {089702} (\bibinfo {year} {2008})%
  \bibAnnoteFile{NoStop}{Schultz2008a}%
\bibitem{Baraff1984}%
  \BibitemOpen
  \bibfield{author}{%
  \bibinfo {author} {\bibfnamefont{G.~A.}\ \bibnamefont{Baraff}}\ and\ \bibinfo
  {author} {\bibfnamefont{M.}~\bibnamefont{{Schl\"uter}}},\ }%
  \bibfield{journal}{%
  \bibinfo {journal} {Phys.\ Rev.\ B}\ }%
  \textbf{\bibinfo {volume} {30}},\ \bibinfo {pages} {1853} (\bibinfo {year}
  {1984})%
  \bibAnnoteFile{NoStop}{Baraff1984}%
\bibitem{Johnson1998a}%
  \BibitemOpen
  \bibfield{author}{%
  \bibinfo {author} {\bibfnamefont{K.~A.}\ \bibnamefont{Johnson}}\ and\
  \bibinfo {author} {\bibfnamefont{N.~W.}\ \bibnamefont{Ashcroft}},\ }%
  \bibfield{journal}{%
  \bibinfo {journal} {Phys.\ Rev.\ B}\ }%
  \textbf{\bibinfo {volume} {58}},\ \bibinfo {pages} {15548} (\bibinfo {year}
  {1998})%
  \bibAnnoteFile{NoStop}{Johnson1998a}%
\bibitem{Monkhorst1976}%
  \BibitemOpen
  \bibfield{author}{%
  \bibinfo {author} {\bibfnamefont{H.~J.}\ \bibnamefont{Monkhorst}}\ and\
  \bibinfo {author} {\bibfnamefont{J.~D.}\ \bibnamefont{Pack}},\ }%
  \bibfield{journal}{%
  \bibinfo {journal} {Phys.\ Rev.\ B}\ }%
  \textbf{\bibinfo {volume} {13}},\ \bibinfo {pages} {5188} (\bibinfo {year}
  {1976})%
  \bibAnnoteFile{NoStop}{Monkhorst1976}%
\bibitem{Becke1990a}%
  \BibitemOpen
  \bibfield{author}{%
  \bibinfo {author} {\bibfnamefont{A.}~\bibnamefont{Becke}}\ and\ \bibinfo
  {author} {\bibfnamefont{K.}~\bibnamefont{Edgecombe}},\ }%
  \bibfield{journal}{%
  \bibinfo {journal} {J.\ Chem.\ Phys.}\ }%
  \textbf{\bibinfo {volume} {92}},\ \bibinfo {pages} {5397} (\bibinfo {year} {1990})%
  \bibAnnoteFile{NoStop}{Becke1990a}%
\bibitem{Wyckoff1963a}%
  \BibitemOpen
  \bibfield{author}{%
  \bibinfo {author} {\bibfnamefont{R.}~\bibnamefont{Wyckoff}},\ }%
  \emph{\bibinfo {title} {Crystal Structures, Second Edition}}\ (\bibinfo
  {publisher} {Wiley, New York},\ \bibinfo {year} {1963})%
  \bibAnnoteFile{NoStop}{Wyckoff1963a}%
\bibitem{Baraff1985a}%
  \BibitemOpen
  \bibfield{author}{%
  \bibinfo {author} {\bibfnamefont{G.~A.}\ \bibnamefont{Baraff}}\ and\ \bibinfo
  {author} {\bibfnamefont{M.}~\bibnamefont{{Schl\"uter}}},\ }%
  \bibfield{journal}{%
  \bibinfo {journal} {Phys.\ Rev.\ Lett.}\ }%
  \textbf{\bibinfo {volume} {55}},\ \bibinfo {pages} {1327} (\bibinfo {year}
  {1985})%
  \bibAnnoteFile{NoStop}{Baraff1985a}%
\bibitem{Northrup1993a}%
  \BibitemOpen
  \bibfield{author}{%
  \bibinfo {author} {\bibfnamefont{J.~E.}\ \bibnamefont{Northrup}}\ and\
  \bibinfo {author} {\bibfnamefont{S.~B.}\ \bibnamefont{Zhang}},\ }%
  \bibfield{journal}{%
  \Doi{10.1103/PhysRevB.47.6791}{\bibinfo {journal} {Phys. Rev. B}}\ }%
  \textbf{\bibinfo {volume} {47}},\ \bibinfo {pages} {6791} (\bibinfo {year} {1993})%
  \bibAnnoteFile{NoStop}{Northrup1993a}%
\bibitem{Gorczyca2002a}%
  \BibitemOpen
  \bibfield{author}{%
  \bibinfo {author} {\bibfnamefont{I.}~\bibnamefont{Gorczyca}}, \bibinfo
  {author} {\bibfnamefont{N.~E.}\ \bibnamefont{Christensen}},\ and\ \bibinfo
  {author} {\bibfnamefont{A.}~\bibnamefont{Svane}},\ }%
  \bibfield{journal}{%
  \Doi{10.1103/PhysRevB.66.075210}{\bibinfo {journal} {Phys. Rev. B}}\ }%
  \textbf{\bibinfo {volume} {66}},\ \bibinfo {pages} {075210} (\bibinfo {year} {2002})%
  \bibAnnoteFile{NoStop}{Gorczyca2002a}%
\bibitem{Janotti2003a}%
  \BibitemOpen
  \bibfield{author}{%
  \bibinfo {author} {\bibfnamefont{A.}~\bibnamefont{Janotti}}, \bibinfo
  {author} {\bibfnamefont{S.-H.}\ \bibnamefont{Wei}}, \bibinfo {author}
  {\bibfnamefont{S.~B.}\ \bibnamefont{Zhang}}, \bibinfo {author}
  {\bibfnamefont{S.}~\bibnamefont{Kurtz}},\ and\ \bibinfo {author}
  {\bibfnamefont{C.~G.}\ \bibnamefont{Van~de Walle}},\ }%
  \bibfield{journal}{%
  \Doi{10.1103/PhysRevB.67.161201}{\bibinfo {journal} {Phys. Rev. B}}\ }%
  \textbf{\bibinfo {volume} {67}},\ \bibinfo {pages} {161201} (\bibinfo {year} {2003})%
  \bibAnnoteFile{NoStop}{Janotti2003a}%
\bibitem{Gebauer2001b}%
  \BibitemOpen
  \bibfield{author}{%
  \bibinfo {author} {\bibfnamefont{J.}~\bibnamefont{Gebauer}}, \bibinfo
  {author} {\bibfnamefont{R.}~\bibnamefont{Zhao}}, \bibinfo {author}
  {\bibfnamefont{P.}~\bibnamefont{Specht}}, \bibinfo {author}
  {\bibfnamefont{E.~R.}\ \bibnamefont{Weber}}, \bibinfo {author}
  {\bibfnamefont{F.}~\bibnamefont{Borner}}, \bibinfo {author}
  {\bibfnamefont{F.}~\bibnamefont{Redmann}},\ and\ \bibinfo {author}
  {\bibfnamefont{R.}~\bibnamefont{Krause-Rehberg}},\ }%
  \bibfield{journal}{%
  \bibinfo {journal} {Appl.\ Phys.\ Lett.}\ }%
  \textbf{\bibinfo {volume} {79}} (\bibinfo {year} {2001})%
  \bibAnnoteFile{NoStop}{Gebauer2001b}%
\bibitem{Yu1991b}%
  \BibitemOpen
  \bibfield{author}{%
  \bibinfo {author} {\bibfnamefont{S.}~\bibnamefont{Yu}}, \bibinfo {author}
  {\bibfnamefont{T.~Y.}\ \bibnamefont{Tan}},\ and\ \bibinfo {author}
  {\bibfnamefont{U.}~\bibnamefont{G\"{o}sele}},\ }%
  \bibfield{journal}{%
  \Doi{10.1063/1.348497}{\bibinfo {journal} {J. Appl.\ Phys.}}\ }%
  \textbf{\bibinfo {volume} {69}},\ \bibinfo {pages} {3547} (\bibinfo {year}
  {1991})%
  \bibAnnoteFile{NoStop}{Yu1991b}%
\end{thebibliography}
\end{document}